\documentclass[aps,prl,amsmath,twocolumn,groupedaddress,letterpaper,floatfix,longbibliography]{revtex4-2}

\usepackage{graphicx,subfigure,epsfig}
\usepackage{array,multirow}
\usepackage{dcolumn}
\usepackage{amssymb,amsmath,amsfonts,mathrsfs}
\usepackage{array}
\usepackage{newtxtext,setspace}
\usepackage{latexsym}
\usepackage{flafter,bm,bbm}
\usepackage{epstopdf,color,multirow}
\usepackage[colorlinks,linkcolor=blue,anchorcolor=blue,urlcolor=blue,citecolor=blue]{hyperref}
\usepackage{footnote}
\usepackage{booktabs}
\usepackage{soul}
\usepackage{cancel}

\makeatletter
\usepackage{dcolumn}
\usepackage{float}
\usepackage{bm}
\usepackage{dsfont}
\usepackage{multirow}
\usepackage{color}

\newcommand{\vket}[1]{\lvert#1\rangle}
\newcommand{\average}[1]{\langle#1\rangle}
\newcommand{\blue}[1]{{\color{blue}{#1}}}

\begin{document}
\title{Quantum Metric Length as a Fundamental Length Scale in Disordered Flat Band Materials}

\author{Chun Wang Chau$^{1,3}$}\thanks{These authors contributed equally to this work}
\author{Tian Xiang$^{1}$}\thanks{These authors contributed equally to this work}
\author{Shuai A. Chen$^{1,2}$}
\email{chsh@ust.hk}
\author{K. T. Law$^{1}$}
\email{phlaw@ust.hk}

\affiliation{1. Department of Physics, Hong Kong University of Science and Technology,
	Clear Water Bay, Hong Kong, China}
\affiliation{2. Max Planck Institute for the Physics of Complex Systems, N\"{o}thnitzer Stra{\ss}e 38, Dresden 01187, Germany}
\affiliation{3. Cavendish Laboratory, Department of Physics, J J Thomson Avenue, Cambridge CB3 0HE, United Kingdom}
\date{\today}

\begin{abstract}
	Our previous understanding of electronic transport in disordered systems was based on the assumption that there is a finite Fermi velocity for the relevant electrons. The Fermi velocity determines important length scales in disordered systems such as the diffusion length and the localization length. However, in disordered systems with vanishing or nearly vanishing Fermi velocity, it is uncertain what determines the important length scales in such systems. In this work, we use the 1D Lieb lattice with isolated flat bands as an example to show that the quantum metric length (QML) is a fundamental length scale in the ballistic, diffusive and localization regimes. The QML is defined through the Bloch state wave functions of the flat bands. In the ballistic regime with short junctions, the QML controls the finite energy transport properties. In the localization regime with long junctions, the localization length is determined by the QML and remarkably, independent of disorder strength over a wide range of disorder strength. We call this unconventional localization regime, the quantum metric localization regime.  In the diffusive regime, we demonstrate that the diffusion coefficient is linearly proportional to the QML via the wave-packet dynamics numerically. Importantly, the numerical results are consistent with the analytical results obtained through the Bethe-Salpeter equation. We conclude that the QML is a fundamentally important length scale governing the properties of disordered flat band materials.
\end{abstract}

\maketitle

\begin{figure}[t]
	\centering 
	\includegraphics[width=1\columnwidth]{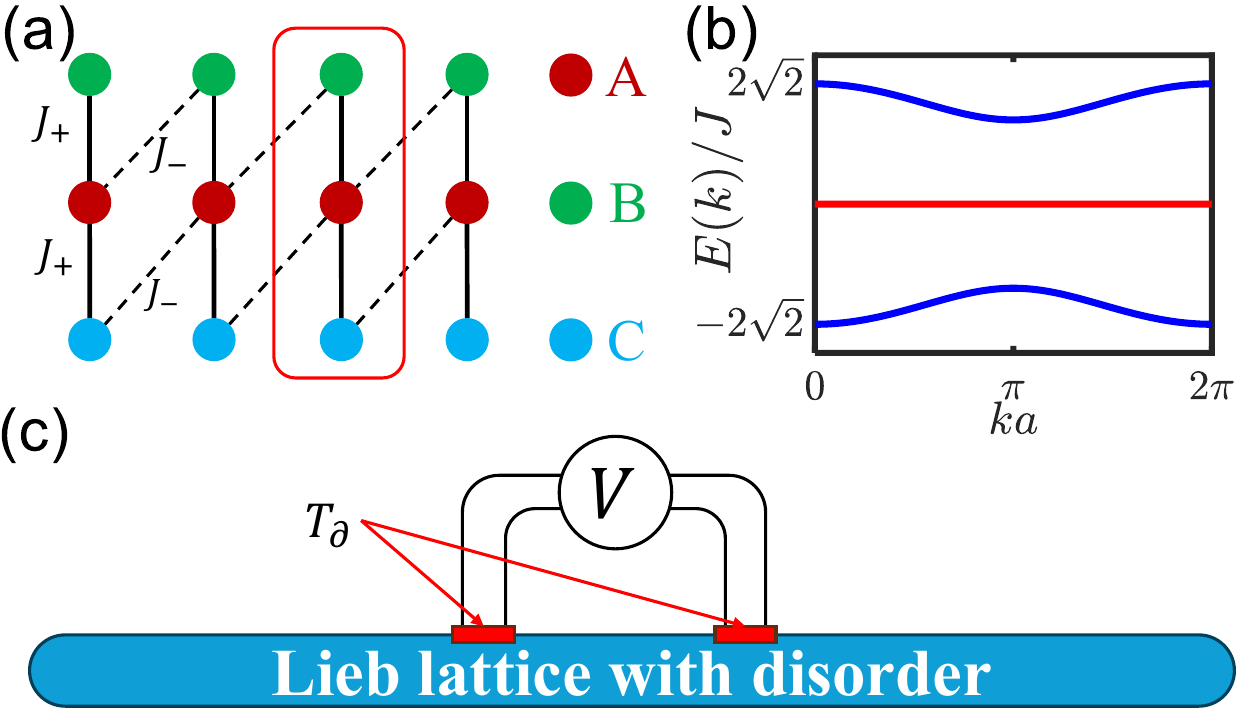}
	\caption{
		(a) A 1D Lieb lattice, which contains three sub-lattices (A, B, and C, respectively) per unit cell. (b)  A schematic band structure of a 1D Lieb lattice with an isolated flat band (red in color). (c) A schematic M/FB/M junction used to study the transport property of a Lieb lattice with disorder. The external leads are connected to the Lieb lattice with coupling strength $T_\partial$, and $V$ denotes the voltage difference of the two leads.
	}
	\label{fig:1} 
\end{figure}


\emph{\color{blue} Introduction---} Flat-band systems, characterized by dispersionless energy bands, have emerged as a fertile platform for exploring diverse quantum phenomena, such as correlated insulating phases~\cite{CIP_1,CIP_2}, superconductivity~\cite{SC_1,SC_2,QM_FB_Josephson,FB-SC_1,FB-SC_2,FB-SC_3,FB-SC_4,FB-SC_5,FB_1,FB_2,FB-SC-6,FB-SC-7}, antiferromagnetism~\cite{FM_1,FM_2,FB-AFM-2}, and excitonic effects~\cite{Ex_1,Ex_2,FB-exciton}. The quantum geometric tensor, which quantifies the phase and amplitude distances between quantum states~\cite{QMT,QM_D_1,QM_D_2,QM-Cond-Mat,Berry_1,Berry_2,Max-Localize-1,Max-Localize-2}, has emerged as a key quantity governing the physical properties of flat-band systems~\cite{QM_SC_1,QM_SC_2,FB_2,Int_1,Dis_1,FB-deg-1,QM-FB-Dis_1,xkxw-1134,FB-transport-2}.  Notably, recent studies suggest a connection between the quantum metric, the real part of the quantum geometric tensor, and the conductivity in flat bands with nontrivial quantum geometry~\cite{QM-FB-Dis_1,Dis_1,FB-deg-1,xkxw-1134,QuantumWeight-1,FB-Trans-1}. 

To highlight an intriguing aspect of flat band disordered systems, we note that in conventional metals, the diffusion coefficient $D= v_{F}l$ is determined by the Fermi velocity $v_{F}$ and the diffusion length $l=v_{F} \tau$ where $\tau$ is the scattering time. In this case, the conductivity is proportional to the diffusion coefficient~\cite{MCMP-Girvin}. For non-interacting flat bands, however, the vanishing Fermi velocity $v_F$ suggests localization behavior~\cite{Kubo-Greenwood-1}. The localization behavior can be altered by  inelastic scatterings~\cite{Dis_1,Dis_2,FB-Nonzero-trans-1}, defects~\cite{QM-FB-Dis_1,QM-FB-Dis_2} and interactions~\cite{Int_1,Int_2,Int_3,FB-Localization-1,FB-Interatcion-1,FB-ABCage-1}, which can potentially induce delocalization. Despite these efforts, some pivotal questions remain. First, as $v_{F}$ goes to zero in flat band materials, what governs the characteristic length scales in the disordered regime? Second, are the diffusive and localized regimes in flat band materials fundamentally the same or different from the corresponding states in systems with large Fermi velocities?

\begin{figure}[t!]
	\includegraphics[width=1\columnwidth]{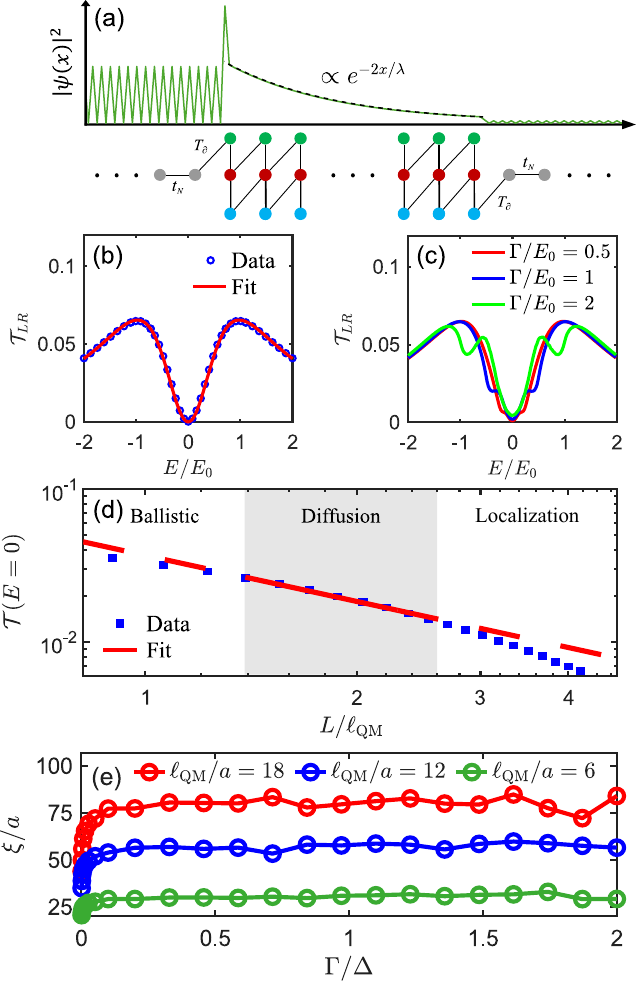}
	\caption{(a) The $|\psi(x)|^{2}=\sum_{\alpha=ABC}|\psi_{\alpha}(x)|^{2}$ of an interface state at the M/FB/M junction where $\psi_{\alpha}(x)$ is the $\alpha$ sub-lattice component of the wave function. Within the Lieb lattice, the probability density of $\psi_{\alpha}(x)$ decays exponentially away from the M/FB interface with a length scale $\lambda=4\ell_\mathrm{QM}$.  
		(b) Transmission probability $\mathcal{T}_{LR}$  for a $L=100$ junction without disorder with the setup of the lower panel of (a). No zero energy transmission is observed, and the transmission at finite energy is mediated by interface states with energy $E_0$. 
		(c) Zero energy transmission arises when disorder is introduced.
		(d) The transmission $\mathcal{T}$ at zero energy $E=0$ for varying junction length $L$ when $\Gamma/E_0=300$ with the setup illustrated in Fig. \ref{fig:1}(c). Three distinct transport regimes are observed where the shaded region indicates the diffusive regime. The diffusive regime is characterized by the $1/L$ scaling, as depicted by the red dashed line.
		(e) The localization length $\xi$ of the localization regime for varying disorder strength $\Gamma$ for three sets of QML $\ell_\mathrm{QM}$. $\xi$ increases in the weak disorder limit, reaching a plateau of $\xi\sim 4\ell_\mathrm{QM}$ when disorder strength is comparable with the band gap $\Delta$.
		The parameters in the calculations are chosen as $t_{N}=1$, $J=1000$, $\delta=0.01$, $T_{\partial}=0.1$, $E_{0}(\delta)=4T_{\partial}^{2}\delta/t_N$ for (b-e).
	}
	\label{fig:2}
\end{figure}

\begin{figure}[t]
	\includegraphics[width=1\columnwidth]{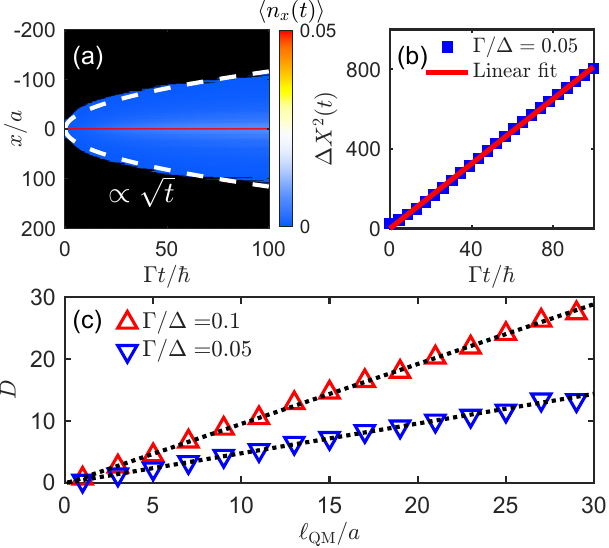} 
	\caption{
		(a) The time evolution of the site occupation $\average{n_x(t)}=\average{\sum_\alpha |\psi_{\alpha}(x,t)|^2}$ for the wave packet $\psi(t)$, where $\psi(t=0)$ is localized and projected to the flat band states.
		(b) The MSD $\Delta X^2(t)$ in (a). The simulation is performed with $\Gamma/\Delta=0.05$. It is clear that $\Delta X^{2}(t)\propto 2Dt$ which indicates a diffusive behavior. (c) The linear fit of $D$ as function of the QML $\ell_{\rm{QM}}$ as predicted from Eq.~\eqref{eq:diffusion}, for disorder strengths $\Gamma/\Delta=0.05$ and $\Gamma/\Delta=0.1$ respectively. The simulation is performed on a 1D Lieb lattice with length $L=401$ in (a-b) and $L=1001$ in (c). All results are obtained by averaging over $500$ disorder configurations.
	}
	\label{fig:3} 
\end{figure}

In this work, we address the above questions by investigating disorder-driven quantum transport in flat-band materials with nontrivial quantum metric. Using the Landauer-B\"uttiker formalism ~\cite{Landauer-1D,Buttiker-1,Buttiker-2,FourTermMeasure-2,FourTermMeasure1,Multiterminal-1,Moire_3}, we study metal/flat-band/metal (M/FB/M) junctions based on Lieb lattice as shown in Fig.~\ref{fig:1}(c) and Fig.~\ref{fig:2}(a). By calculating the transmission probability between the two leads in the presence of disorder, we unveil the disorder-induced transport properties in isolated flat bands with quantum metric. First, in the clean limit, transmission at a finite energy is mediated by the interface states whose localization length is set by the quantum metric length (QML) as defined in Eq.~\eqref{eq:QML}. Second, disorders enable bulk-state transmission at zero energy as shown in Fig.~\ref{fig:2}(c). Thirdly, a highly conventional localization behavior is demonstrated for samples much longer than the QML and the localization length is determined by the QML as depicted in Fig.~\ref{fig:2}(e). Remarkably, the localization length is a constant, set by the QML, for a wide range of disorder strength. This is in sharp contrast to the Anderson localization regime in which the localization length decreases as the disorder strength increases. We call this regime, the quantum metric localization regime. Fourthly, we observe a diffusive transport regime when the sample size is comparable with the QML as shown in Fig.~\ref{fig:2}(d). Furthermore, via wave-packet dynamics calculations, we show that the diffusion coefficient is linearly proportional to the QML as shown in Fig.~\ref{fig:3}. Interestingly, the diffusion coefficient is linearly proportional to disorder strength which contradicts the understandings from conventional theories. Importantly, we show that the numerical results are consistent with the results from the Bethe-Salpeter equation as summerized in Eq.~\eqref{eq:diffusion}. This work shows that the QML is the fundamental length scale in the flat band material in the ballistic, diffusive and localization regimes. Moreover, the diffusive and the localization regimes have anomalous dependence on the disorder strength. 

\emph{\color{blue} Quantum Metric Length of Lieb Lattice Model ---}\label{Lieb} The M/FB/M junction for the study of disorder effects is constructed by connecting two metallic leads to a Lieb lattice as depicted in Fig.~\ref{fig:1}(a) and Fig.~\ref{fig:2}(a). In this section, we introduce the Lieb lattice with flat bands at the Fermi energy. First of all, each unit cell of the Lieb lattice hosts three sub-lattices (A, B, and C, respectively), with corresponding annihilation operators labeled as $\hat{a}_{x}$, $\hat{b}_{x}$, and $\hat{c}_{x}$, respectively. The Hamiltonian for a Lieb lattice reads $\hat{H}_{\text{Lieb}} = \sum_x \hat h_x$ with
\begin{align}
	\!\!\!\! \hat h_{x} &=J_{+}(\hat{b}_{x}^{\dagger}\hat{a}_{x}+\hat{c}_{x}^{\dagger}\hat{a}_{x})+J_{-}(\hat{a}_{x}^{\dagger}\hat{b}_{x+1}+\hat{c}_{x}^{\dagger}\hat{a}_{x+1})+h.c.,
\end{align}
where $J_{\pm} = J(1 \pm \delta)$ are the intra- and inter-cell hopping amplitudes respectively, and $x$ labels the lattice site.  The Lieb lattice features an isolated flat band which is separated from the two dispersive bands by a gap $\Delta = 2\sqrt{2}J\delta$ as illustrated in Fig.~\ref{fig:1}(b).
The quantum metric for a Bloch state $\vket{u(k)}$ of the flat band with momentum $k$ is defined as $\mathcal{G}(k) = \langle\partial_{k}u(k)|(1-|u(k)\rangle\langle u(k)|)|\partial_{k}u(k)\rangle$. The Brillouin-zone average of the quantum metric of the flat band is given by
\begin{equation}\label{eq:qm}
	\overline{\mathcal{G}}=\frac{a}{2\pi}\int_{-\pi/a}^{\pi/a}\mathcal{G}(k)dk=\frac{a^2(1-\delta)^2}{8\delta},
\end{equation}
which has the dimension of length squared. The QML  $\ell_{\rm{QM}}$ is defined as:
\begin{equation}\label{eq:QML}
	\ell_{\rm{QM}}=\overline{\mathcal{G}}/a,
\end{equation}
where $a$ is the lattice constant, and we set $a=1$ throughout this work. Previous studies~\cite{QM_SC_1,QM_SC_2,Maj_1,FB-JJ-3} have shown that the QML determines the superconducting coherence length of the flat band superconductors~\cite{QM_SC_1, QM_SC_2} and the localization length of topological bound states~\cite{Maj_1}.
The question is, what is the role of the QML in disordered flat band systems when $v_{F}$ goes to zero and important length scales such as the conventional diffusion length and the localization length vanish?

.

\emph{\color{blue} Disorder-free case ---} Before investigating the disordered cases, we first show the importance of the QML $\ell_{\rm{QM}}$ without disorder. As depicted in Fig.~\ref{fig:2}(a) (lower panel), two metallic leads with bandwidth $2t_N$ are attached to the Lieb lattice and the Fermi energy is set at the flat band energy. The coupling amplitude between the lead and the Lieb lattice is denoted as $T_{\partial}$. The transmission probability from the left lead to the right lead $\mathcal{T}_{LR}$ is calculated as $\mathcal{T}_{LR} = {\rm Tr}[\Gamma_{L}G^{R}\Gamma_{R}(G^{R})^{\dagger}]$, where $\Gamma_{L/R}$ is the linewidth functions induced by the left and right leads respectively, and $G^{R}$ is the retarded Green's function of the lattice Hamiltonian $\hat{H}_{\text{Lieb}}$~\cite{supple,Multiterminal-1,FourTermMeasure1}.  

As shown in Fig.~\ref{fig:2}(b) for $\mathcal{T}_{LR}$ versus $E$, where $E$ is the energy difference between the leads' Fermi energy and the flat band  energy of Lieb lattice, there is no zero energy transmission in the absence of disorders. This is consistent with the intuition that the conductance is zero when the Fermi velocity vanishes~\cite{Kubo-Greenwood-1}. Interestingly, when the two metallic leads are coupled to the Lieb lattice, two interface states located at the M/FB and the FB/M interfaces are created. Importantly, the interface states have a decay length $\lambda$ set by the QML as $\lambda = 4\ell_{\rm{QM}}$ ~\cite{supple}. The probability density of an interface state $\psi(x)$ at M/FB interface is shown in Fig.~\ref{fig:2}(a) where $x$ is the site index of the Lieb lattice. When the junction length is shorter or comparable with $\lambda$, the two left and right interface states hybridize to form a finite energy resonant state for electrons to tunnel through the junction. As a result, there are tunneling peaks at finite energy $E_0=4T^2_\partial\delta/t_N$ as depicted in Fig.~\ref{fig:2}(b). In the weak coupling limit, we can show analytically by two different methods~\cite{supple} that the tunneling probability at energy $E$ is given by
\begin{equation}
	\mathcal{T}_{LR}(E) \approx \frac{16E^{2}E_{0}^{2}}{(E^{2}+E_{0}^{2})^{2}}e^{-\frac{2L}{\lambda}}.
\end{equation}
The analytic expression of $\mathcal{T}_{LR}(E)$ (red curve in Fig.~\ref{fig:2}(b)) matches the numerical results nearly perfectly. It is important to note that the $\mathcal{T}_{LR}(E)$ is controlled by the length scale $\lambda = 4\ell_{\rm{QM}}$ which clearly demonstrates the importance of the QML in the flat band materials in the clean limit.

\emph{\color{blue} Disorder effects and three transport regimes---} To study the effects of disorder, we introduce Anderson-type onsite disorder to the Lieb lattice:
\begin{equation}
	\hat{H}_{\text{dis}}=\sum_{x}w_{x}(\hat{a}_{x}^{\dagger}\hat{a}_{x}+\hat{b}_{x}^{\dagger}\hat{b}_{x}+\hat{c}_{x}^{\dagger}\hat{c}_{x}),\label{eq:dis}
\end{equation}
where the onsite potential $w_{x}$ is independently and uniformly distributed in the range $[-\Gamma/2, \Gamma/2]$.

In the presence of disorder, the tunneling peaks near energy $E_0$ are smeared out as shown in Fig.~\ref{fig:2}(c). Importantly, finite transmission probability emerges at zero energy such that we have disorder enhanced electronic transport. Each data point in Fig.~\ref{fig:2}(c) for $\mathcal{T}_{LR}(E)$ is the average of results from $10000$ disorder configurations. As the zero energy transport is away from the resonant state with energy $E_0$, it is likely that the zero energy transmission is caused by the bulk transport as discussed below.

To understand the bulk state transport in the presence of disorder, we examine the dependence of the transmission on the sample size. Importantly, to minimize the contribution from the interface states as shown in Fig.~\ref{fig:2}(a), we connect the two metallic leads to the bulk of the Lieb lattice model as illustrated in Fig.~\ref{fig:1}(c). With the setup in Fig.~\ref{fig:1}(c), the zero energy transmission probability $\mathcal{T}(E=0)$ as a function of junction length $L$ is computed and the results are shown in Fig.~\ref{fig:2}(d). Interestingly, three distinct transport regimes are observed depending on $L$. We identify the three transport regimes as the ballistic, diffusive and the localization regime, respectively.

For short junctions with $L < \ell_{\rm QM}$, the transmission decreases linearly with length, following the relation $\mathcal{T} \propto 1 - L/\xi$~\citep{Q1D-Diffu}. We identify this regime as the ballistic regime \cite{Dis-Cond-1} ($L \ll \xi$) where $\xi$ is in the order of the QML as shown in Fig.~\ref{fig:2}(d). On the other hand, for sufficiently long junctions ($L \gg \ell_{\rm QM}$), the system transitions to the localized regime, where the transmission decays exponentially with $L$ such that $\mathcal{T} \propto e^{-L/\xi}$~\cite{And-Loc-1,FBL-AL-1D-Exper,FB-Loc-5}. The localization lengths with different disorder strengths for long junctions are extracted through the calculations of $\mathcal{T} \propto e^{-L/\xi}$ and the results are shown in  Fig.~\ref{fig:2}(e). It is interesting that the localization length \emph{increases} as the disorder strength  \emph{increases} when $\Gamma$ is small compared to the band gap $\Delta$ of the Lieb lattice in this weak disorder regime. This is in sharp contrast to the Anderson localization scenario when the localization length decreases as the disorder strength increases. Remarkably, as shown in Fig.~\ref{fig:2}(e), for a wide range of disorder strength in the order of $\Delta$, the localization length is approximately close to $\lambda = 4\ell_{\rm{QM}}$. We call this regime, the quantum metric localization regime where the localization length is determined by the QML and independent of disorder strength. This shows again that the QML is the important length scale in this unconventional localization regime. 

When the junction length is comparable to the QML  $\ell_{\rm QM}$ such that $L  \approx \ell_{\rm QM}$ [the shaded region in Fig.~\ref{fig:2}(d)], we have $\mathcal{T} \propto L^{-1}$ such that the Ohmic relation is satisfied ~\cite{Diff-Len-1}. The Ohmic relation is clearly shown by a straight line with slope $-1$ in the log-log plot of Fig.~\ref{fig:2}(d). In this regime, disorders disrupt the perfect interference that caused the formation of the exact flat band, allowing the electrons to propagate as in conventional disordered systems~\citep{Review-QuantTrans}. 
However, in conventional theories, the diffusion length is expected to be zero as the Fermi velocity is zero for a flat band. The question is, what is the length scale governing this diffusive transport regime? This question will be answered in the next section.


\emph{\color{blue} Wave packet dynamics calculations for the diffusion coefficient.---} To further understand the properties of the diffusive transport regime, we study the short-time evolution of wavefunctions through the wave packet dynamics. In the wave packet dynamics, the diffusion coefficient can be extracted from the time dependent mean square displacement (MSD) $\Delta X^2(t)$ as $D=\frac{1}{2}\frac{d(\Delta X^2)}{dt}$, which can be calculated through ~\cite{Diffusion-MSD-1,Diffusion-MSD-2,FB-WavePacket-1,FB-Photon-1,FB-SC-Qbit-Array}:
\begin{equation}\label{eq:MSD}
	\Delta X^2(t)=\sum_{x=-L/2}^{L/2}x^2 \average{n_x(t)}-\left(\sum_{x=-L/2}^{L/2}x\, \average{n_x(t)}\right)^2,
\end{equation}
where $\average{\cdot}$ denotes the disorder average and $n_x(t)=\sum_\alpha |\psi_{\alpha}(x,t)|^2$ is the occupation number at site $x$ at time $t$, and $\alpha \in (A,B,C)$ is the sublattice index. The MSD $\Delta X^2(t)$ measures how far the wave packet can spread over time. In particular, if the wave packet evolves diffusively, the MSD will grow linearly with time as $\Delta X^2(t)=2Dt$ ~\cite{Diffusion-MSD-1,Diffusion-MSD-2,FB-WavePacket-1,FB-Photon-1}.

In Fig.~\ref{fig:3}(a), we start with a localized wave packet and project the wave function to the flat-band states \cite{DL-Diffusion, supple} to obtain an initial wavefunction $\vket{\psi_0}$. Then, the wave function evolves under the time-evolution operator such that $\vket{\psi(t)}=e^{-i\hat{H}t/\hbar}\vket{\psi_0}$, where $\hat{H}$ contains Lieb lattice as well as the disorder term $\hat{H}_{\text{dis}}$. It is clear from Fig.~\ref{fig:3}(b) that the $\Delta X^2(t)$ scales linearly as a function of time, indicating that an initially localized wave packet diffuses via a random walk process~\cite{Diffusion-MSD-1,Diffusion-MSD-3,Diffusion-Noise-1}. If the initial localized state is projected to the dispersive bands, the $\Delta X^2(t)$ of the wave packet scales as $t^2$ instead as shown in the Supplemental Material~\cite{supple}, indicating the ballistic transport behavior in this weak disorder regime.

In Fig.~\ref{fig:3}(c), we further study the diffusive regime by tuning the QML, for two different disorder strengths. In particular, as the QML increases, the diffusion coefficient extracted from wave packet dynamics increases linearly accordingly.  The results in Fig.~\ref{fig:3} strongly indicate that the diffusion coefficient is linearly proportional to the QML in the diffusive regime. It is also clear that the diffusion coefficient increases as the disorder strength increases which is highly unconventional.


\emph{\color{blue} Solving the Bethe-Salpeter equation for the diffusion coefficient.---} 
To have an analytical understanding of the observation that the diffusion coefficient is linearly proportional to the QML, we apply the diagrammatic techniques to obtain the diffusion coefficient through the Bethe-Salpeter equation~\citep{Book-Meso,Book-Bruus}. Employing the ladder approximation, the impurity vertex encapsulating the diffusion regime $\Pi(\omega,q)$ can be solved self-consistently:
\begin{equation}\label{eq:BetheSalpeter}
	\Pi(\omega,q)=\Pi_{0}(\omega,q)+P_{0}(\omega,q)\Pi_{0}(\omega,q)\Pi(\omega,q).
\end{equation}
Here, $P_{0}(\omega,q)$ is the Fourier transform of $P_{0}(t,x,x')$ which is the probability to propagate from position $x$ to $x'$ in time $t$ without collisions. On the other hand, $\Pi_{0}(\omega,q)$ is the bare impurity vertex, describing a single scattering process due to disorder.
For a conventional dispersive band, the impurity vertex takes the form $\Pi(\omega,q)\propto 1/(-i\omega+Dq^2)$~\citep{Book-Meso,Book-Bruus,Book-allen}, with  $D=v_Fl$ identified as the diffusion coefficient.

For the flat band case, the bare impurity vertex takes the form $\Pi_{0}(\omega,q)=\int\frac{dk}{2\pi}|\langle u(k)|u(k+q)\rangle|^{2}$, which reduces to $\Pi_{0}(\omega,q)\approx\Gamma^2(1-q^{2}\overline{\mathcal{G}})$ in the small $q$ limit. The diffusion probability without collisions is given by $P_{0}(\omega,q)=\overline{G}(E)\overline{G}(E+\omega)$, where $\overline{G}(E)\sim 2/(E+i\Gamma/\sqrt{3})$ is the disorder-averaged Green's function for $E\ll\Gamma$. In this case, solving Eq.~\eqref{eq:BetheSalpeter} gives rise to the full impurity vertex $\Pi(\omega,q) \propto \frac{ 1 }{ -i\omega +  C\Gamma \overline{\mathcal{G}} q^2}$. The flat band impurity vertex \cite{supple} has the form as the dispersive band case and we can identify the diffusion coefficient as: 

\begin{equation}\label{eq:diffusion}
	D=C\times\Gamma\, \ell_{\mathrm{QM}}a.
\end{equation}
The results reveal that for the flat band system, the diffusion coefficient scales linearly with the QML $\ell_{\mathrm{QM}}$. 
Incredibly, Eq.~\eqref{eq:diffusion} is in agreement with the numerical results depicted in Fig.~\ref{fig:3}(c). Eq.~\eqref{eq:diffusion} is significant as it generally applies to 1D systems with isolated flat bands and is not limited to the Lieb lattice chosen in the numerical studies. The proportionality constant $C\approx 0.337$ has been extracted numerically for the 1D Lieb lattice using data displayed in Fig.~\ref{fig:3}(c).


\emph{\blue{Discussion and conclusion.---}}  In this work, using the 1D Lieb lattice, we demonstrate that the QML is the fundamental length scale governing the transport properties of flat band materials in the ballistic, diffusive as well as the localization regimes. Importantly, a highly unconventional quantum metric localization regime was identified in which the localization length is determined by the QML and independent of the disorder strength. Another unconventional diffusive regime was identified in which the diffusion constant is linearly proportional to the QML as well as the disorder strength. Using the Bethe-Salpeter equation, the numerical results in the diffusion regime were verified. The analytical result goes beyond the Lieb lattice and it generally applies to 1D lattices with isolated flat bands. In the future, more works will be needed to further establish the importance of QML in disordered flat band materials with other models~\cite{Flat-Tasaki-1}. 

In realistic materials such as in twisted bilayer graphene ~\cite{Moire_1,Moire_2,Dis-MATBG-1}, the energy bands are generally dispersive. We expect that our results still apply when the QML is longer than the length scales introduced by the Fermi velocity. Furthermore, one important and natural extension of the current work is to generalize the study to two-dimensional flat band materials. Due to the absence of Fermi velocity related length scales, we believe that the QML still plays an important role in two-dimensional flat band systems. As the QML can also be defined in electronic, photonic and phononic systems, our results are can be easily extended to photonic and phononic systems as well~\citep{QDA, Op_Lat_1,Kagome_1,CLS-1,CLS-2}.

\begin{acknowledgments} 
	We thank Patrick A. Lee, Tai-Kai Ng, Carlo Beenakker, Roderich Moessner, Akito Daido, Sen Mu, and Haijing Zhang for their valuable discussions. K. T. L. acknowledges the support of the Ministry of Science and Technology, China, and the Hong Kong Research Grants Council through Grants No. MOST23SC01-A, No. RFS2021-6S03, No. C6025-19G, No. AoE/P-701/20, No. 16310520, No. 16310219, No. 16307622, and No. 16309223. K.T.L. is also supported by the New Cornerstone Investigator Program. 
\end{acknowledgments}


\begin{thebibliography}{85}%
	\makeatletter
	\providecommand \@ifxundefined [1]{%
		\@ifx{#1\undefined}
	}%
	\providecommand \@ifnum [1]{%
		\ifnum #1\expandafter \@firstoftwo
		\else \expandafter \@secondoftwo
		\fi
	}%
	\providecommand \@ifx [1]{%
		\ifx #1\expandafter \@firstoftwo
		\else \expandafter \@secondoftwo
		\fi
	}%
	\providecommand \natexlab [1]{#1}%
	\providecommand \enquote  [1]{``#1''}%
	\providecommand \bibnamefont  [1]{#1}%
	\providecommand \bibfnamefont [1]{#1}%
	\providecommand \citenamefont [1]{#1}%
	\providecommand \href@noop [0]{\@secondoftwo}%
	\providecommand \href [0]{\begingroup \@sanitize@url \@href}%
	\providecommand \@href[1]{\@@startlink{#1}\@@href}%
	\providecommand \@@href[1]{\endgroup#1\@@endlink}%
	\providecommand \@sanitize@url [0]{\catcode `\\12\catcode `\$12\catcode
		`\&12\catcode `\#12\catcode `\^12\catcode `\_12\catcode `\%12\relax}%
	\providecommand \@@startlink[1]{}%
	\providecommand \@@endlink[0]{}%
	\providecommand \url  [0]{\begingroup\@sanitize@url \@url }%
	\providecommand \@url [1]{\endgroup\@href {#1}{\urlprefix }}%
	\providecommand \urlprefix  [0]{URL }%
	\providecommand \Eprint [0]{\href }%
	\providecommand \doibase [0]{http://dx.doi.org/}%
	\providecommand \selectlanguage [0]{\@gobble}%
	\providecommand \bibinfo  [0]{\@secondoftwo}%
	\providecommand \bibfield  [0]{\@secondoftwo}%
	\providecommand \translation [1]{[#1]}%
	\providecommand \BibitemOpen [0]{}%
	\providecommand \bibitemStop [0]{}%
	\providecommand \bibitemNoStop [0]{.\EOS\space}%
	\providecommand \EOS [0]{\spacefactor3000\relax}%
	\providecommand \BibitemShut  [1]{\csname bibitem#1\endcsname}%
	\let\auto@bib@innerbib\@empty
	\bibitem [{\citenamefont {{Cao}}\ \emph
		{et~al.}(2018{\natexlab{a}})\citenamefont {{Cao}}, \citenamefont {{Fatemi}},
		\citenamefont {{Demir}}, \citenamefont {{Fang}}, \citenamefont {{Tomarken}},
		\citenamefont {{Luo}}, \citenamefont {{Sanchez-Yamagishi}}, \citenamefont
		{{Watanabe}}, \citenamefont {{Taniguchi}}, \citenamefont {{Kaxiras}},
		\citenamefont {{Ashoori}},\ and\ \citenamefont {{Jarillo-Herrero}}}]{CIP_1}%
	\BibitemOpen
	\bibfield  {author} {\bibinfo {author} {\bibfnamefont {Yuan}\ \bibnamefont
			{{Cao}}}, \bibinfo {author} {\bibfnamefont {Valla}\ \bibnamefont {{Fatemi}}},
		\bibinfo {author} {\bibfnamefont {Ahmet}\ \bibnamefont {{Demir}}}, \bibinfo
		{author} {\bibfnamefont {Shiang}\ \bibnamefont {{Fang}}}, \bibinfo {author}
		{\bibfnamefont {Spencer~L.}\ \bibnamefont {{Tomarken}}}, \bibinfo {author}
		{\bibfnamefont {Jason~Y.}\ \bibnamefont {{Luo}}}, \bibinfo {author}
		{\bibfnamefont {Javier~D.}\ \bibnamefont {{Sanchez-Yamagishi}}}, \bibinfo
		{author} {\bibfnamefont {Kenji}\ \bibnamefont {{Watanabe}}}, \bibinfo
		{author} {\bibfnamefont {Takashi}\ \bibnamefont {{Taniguchi}}}, \bibinfo
		{author} {\bibfnamefont {Efthimios}\ \bibnamefont {{Kaxiras}}}, \bibinfo
		{author} {\bibfnamefont {Ray~C.}\ \bibnamefont {{Ashoori}}}, \ and\ \bibinfo
		{author} {\bibfnamefont {Pablo}\ \bibnamefont {{Jarillo-Herrero}}},\
	}\bibfield  {title} {\enquote {\bibinfo {title} {{Correlated insulator
					behaviour at half-filling in magic-angle graphene superlattices}},}\ }\href
	{\doibase 10.1038/nature26154} {\bibfield  {journal} {\bibinfo  {journal}
			{\nat}\ }\textbf {\bibinfo {volume} {556}},\ \bibinfo {pages} {80--84}
		(\bibinfo {year} {2018}{\natexlab{a}})},\ \Eprint
	{http://arxiv.org/abs/1802.00553} {arXiv:1802.00553 [cond-mat.mes-hall]}
	\BibitemShut {NoStop}%
	\bibitem [{\citenamefont {{Regan}}\ \emph {et~al.}(2020)\citenamefont
		{{Regan}}, \citenamefont {{Wang}}, \citenamefont {{Jin}}, \citenamefont
		{{Bakti Utama}}, \citenamefont {{Gao}}, \citenamefont {{Wei}}, \citenamefont
		{{Zhao}}, \citenamefont {{Zhao}}, \citenamefont {{Zhang}}, \citenamefont
		{{Yumigeta}}, \citenamefont {{Blei}}, \citenamefont {{Carlstr{\"o}m}},
		\citenamefont {{Watanabe}}, \citenamefont {{Taniguchi}}, \citenamefont
		{{Tongay}}, \citenamefont {{Crommie}}, \citenamefont {{Zettl}},\ and\
		\citenamefont {{Wang}}}]{CIP_2}%
	\BibitemOpen
	\bibfield  {author} {\bibinfo {author} {\bibfnamefont {Emma~C.}\ \bibnamefont
			{{Regan}}}, \bibinfo {author} {\bibfnamefont {Danqing}\ \bibnamefont
			{{Wang}}}, \bibinfo {author} {\bibfnamefont {Chenhao}\ \bibnamefont {{Jin}}},
		\bibinfo {author} {\bibfnamefont {M.~Iqbal}\ \bibnamefont {{Bakti Utama}}},
		\bibinfo {author} {\bibfnamefont {Beini}\ \bibnamefont {{Gao}}}, \bibinfo
		{author} {\bibfnamefont {Xin}\ \bibnamefont {{Wei}}}, \bibinfo {author}
		{\bibfnamefont {Sihan}\ \bibnamefont {{Zhao}}}, \bibinfo {author}
		{\bibfnamefont {Wenyu}\ \bibnamefont {{Zhao}}}, \bibinfo {author}
		{\bibfnamefont {Zuocheng}\ \bibnamefont {{Zhang}}}, \bibinfo {author}
		{\bibfnamefont {Kentaro}\ \bibnamefont {{Yumigeta}}}, \bibinfo {author}
		{\bibfnamefont {Mark}\ \bibnamefont {{Blei}}}, \bibinfo {author}
		{\bibfnamefont {Johan~D.}\ \bibnamefont {{Carlstr{\"o}m}}}, \bibinfo {author}
		{\bibfnamefont {Kenji}\ \bibnamefont {{Watanabe}}}, \bibinfo {author}
		{\bibfnamefont {Takashi}\ \bibnamefont {{Taniguchi}}}, \bibinfo {author}
		{\bibfnamefont {Sefaattin}\ \bibnamefont {{Tongay}}}, \bibinfo {author}
		{\bibfnamefont {Michael}\ \bibnamefont {{Crommie}}}, \bibinfo {author}
		{\bibfnamefont {Alex}\ \bibnamefont {{Zettl}}}, \ and\ \bibinfo {author}
		{\bibfnamefont {Feng}\ \bibnamefont {{Wang}}},\ }\bibfield  {title} {\enquote
		{\bibinfo {title} {{Mott and generalized Wigner crystal states in
					WSe$_{2}$/WS$_{2}$ moir{\'e} superlattices}},}\ }\href {\doibase
		10.1038/s41586-020-2092-4} {\bibfield  {journal} {\bibinfo  {journal} {\nat}\
		}\textbf {\bibinfo {volume} {579}},\ \bibinfo {pages} {359--363} (\bibinfo
		{year} {2020})},\ \Eprint {http://arxiv.org/abs/1910.09047} {arXiv:1910.09047
		[cond-mat.mes-hall]} \BibitemShut {NoStop}%
	\bibitem [{\citenamefont {{Cao}}\ \emph
		{et~al.}(2018{\natexlab{b}})\citenamefont {{Cao}}, \citenamefont {{Fatemi}},
		\citenamefont {{Fang}}, \citenamefont {{Watanabe}}, \citenamefont
		{{Taniguchi}}, \citenamefont {{Kaxiras}},\ and\ \citenamefont
		{{Jarillo-Herrero}}}]{SC_1}%
	\BibitemOpen
	\bibfield  {author} {\bibinfo {author} {\bibfnamefont {Yuan}\ \bibnamefont
			{{Cao}}}, \bibinfo {author} {\bibfnamefont {Valla}\ \bibnamefont {{Fatemi}}},
		\bibinfo {author} {\bibfnamefont {Shiang}\ \bibnamefont {{Fang}}}, \bibinfo
		{author} {\bibfnamefont {Kenji}\ \bibnamefont {{Watanabe}}}, \bibinfo
		{author} {\bibfnamefont {Takashi}\ \bibnamefont {{Taniguchi}}}, \bibinfo
		{author} {\bibfnamefont {Efthimios}\ \bibnamefont {{Kaxiras}}}, \ and\
		\bibinfo {author} {\bibfnamefont {Pablo}\ \bibnamefont {{Jarillo-Herrero}}},\
	}\bibfield  {title} {\enquote {\bibinfo {title} {{Unconventional
					superconductivity in magic-angle graphene superlattices}},}\ }\href {\doibase
		10.1038/nature26160} {\bibfield  {journal} {\bibinfo  {journal} {\nat}\
		}\textbf {\bibinfo {volume} {556}},\ \bibinfo {pages} {43--50} (\bibinfo
		{year} {2018}{\natexlab{b}})},\ \Eprint {http://arxiv.org/abs/1803.02342}
	{arXiv:1803.02342 [cond-mat.mes-hall]} \BibitemShut {NoStop}%
	\bibitem [{\citenamefont {{Lu}}\ \emph {et~al.}(2019)\citenamefont {{Lu}},
		\citenamefont {{Stepanov}}, \citenamefont {{Yang}}, \citenamefont {{Xie}},
		\citenamefont {{Aamir}}, \citenamefont {{Das}}, \citenamefont {{Urgell}},
		\citenamefont {{Watanabe}}, \citenamefont {{Taniguchi}}, \citenamefont
		{{Zhang}}, \citenamefont {{Bachtold}}, \citenamefont {{MacDonald}},\ and\
		\citenamefont {{Efetov}}}]{SC_2}%
	\BibitemOpen
	\bibfield  {author} {\bibinfo {author} {\bibfnamefont {Xiaobo}\ \bibnamefont
			{{Lu}}}, \bibinfo {author} {\bibfnamefont {Petr}\ \bibnamefont {{Stepanov}}},
		\bibinfo {author} {\bibfnamefont {Wei}\ \bibnamefont {{Yang}}}, \bibinfo
		{author} {\bibfnamefont {Ming}\ \bibnamefont {{Xie}}}, \bibinfo {author}
		{\bibfnamefont {Mohammed~Ali}\ \bibnamefont {{Aamir}}}, \bibinfo {author}
		{\bibfnamefont {Ipsita}\ \bibnamefont {{Das}}}, \bibinfo {author}
		{\bibfnamefont {Carles}\ \bibnamefont {{Urgell}}}, \bibinfo {author}
		{\bibfnamefont {Kenji}\ \bibnamefont {{Watanabe}}}, \bibinfo {author}
		{\bibfnamefont {Takashi}\ \bibnamefont {{Taniguchi}}}, \bibinfo {author}
		{\bibfnamefont {Guangyu}\ \bibnamefont {{Zhang}}}, \bibinfo {author}
		{\bibfnamefont {Adrian}\ \bibnamefont {{Bachtold}}}, \bibinfo {author}
		{\bibfnamefont {Allan~H.}\ \bibnamefont {{MacDonald}}}, \ and\ \bibinfo
		{author} {\bibfnamefont {Dmitri~K.}\ \bibnamefont {{Efetov}}},\ }\bibfield
	{title} {\enquote {\bibinfo {title} {{Superconductors, orbital magnets and
					correlated states in magic-angle bilayer graphene}},}\ }\href {\doibase
		10.1038/s41586-019-1695-0} {\bibfield  {journal} {\bibinfo  {journal} {\nat}\
		}\textbf {\bibinfo {volume} {574}},\ \bibinfo {pages} {653--657} (\bibinfo
		{year} {2019})},\ \Eprint {http://arxiv.org/abs/1903.06513} {arXiv:1903.06513
		[cond-mat.str-el]} \BibitemShut {NoStop}%
	\bibitem [{\citenamefont {{Pyykk{\"o}nen}}\ \emph {et~al.}(2021)\citenamefont
		{{Pyykk{\"o}nen}}, \citenamefont {{Peotta}}, \citenamefont {{Fabritius}},
		\citenamefont {{Mohan}}, \citenamefont {{Esslinger}},\ and\ \citenamefont
		{{T{\"o}rm{\"a}}}}]{QM_FB_Josephson}%
	\BibitemOpen
	\bibfield  {author} {\bibinfo {author} {\bibfnamefont {Ville A.~J.}\
			\bibnamefont {{Pyykk{\"o}nen}}}, \bibinfo {author} {\bibfnamefont
			{Sebastiano}\ \bibnamefont {{Peotta}}}, \bibinfo {author} {\bibfnamefont
			{Philipp}\ \bibnamefont {{Fabritius}}}, \bibinfo {author} {\bibfnamefont
			{Jeffrey}\ \bibnamefont {{Mohan}}}, \bibinfo {author} {\bibfnamefont
			{Tilman}\ \bibnamefont {{Esslinger}}}, \ and\ \bibinfo {author}
		{\bibfnamefont {P{\"a}ivi}\ \bibnamefont {{T{\"o}rm{\"a}}}},\ }\bibfield
	{title} {\enquote {\bibinfo {title} {{Flat-band transport and Josephson
					effect through a finite-size sawtooth lattice}},}\ }\href {\doibase
		10.1103/PhysRevB.103.144519} {\bibfield  {journal} {\bibinfo  {journal}
			{\prb}\ }\textbf {\bibinfo {volume} {103}},\ \bibinfo {eid} {144519}
		(\bibinfo {year} {2021})},\ \Eprint {http://arxiv.org/abs/2101.04460}
	{arXiv:2101.04460 [cond-mat.quant-gas]} \BibitemShut {NoStop}%
	\bibitem [{\citenamefont {{Pyykk{\"o}nen}}\ \emph {et~al.}(2023)\citenamefont
		{{Pyykk{\"o}nen}}, \citenamefont {{Peotta}},\ and\ \citenamefont
		{{T{\"o}rm{\"a}}}}]{FB-SC_1}%
	\BibitemOpen
	\bibfield  {author} {\bibinfo {author} {\bibfnamefont {Ville A.~J.}\
			\bibnamefont {{Pyykk{\"o}nen}}}, \bibinfo {author} {\bibfnamefont
			{Sebastiano}\ \bibnamefont {{Peotta}}}, \ and\ \bibinfo {author}
		{\bibfnamefont {P{\"a}ivi}\ \bibnamefont {{T{\"o}rm{\"a}}}},\ }\bibfield
	{title} {\enquote {\bibinfo {title} {{Suppression of Nonequilibrium
					Quasiparticle Transport in Flat-Band Superconductors}},}\ }\href {\doibase
		10.1103/PhysRevLett.130.216003} {\bibfield  {journal} {\bibinfo  {journal}
			{\prl}\ }\textbf {\bibinfo {volume} {130}},\ \bibinfo {eid} {216003}
		(\bibinfo {year} {2023})},\ \Eprint {http://arxiv.org/abs/2211.09483}
	{arXiv:2211.09483 [cond-mat.supr-con]} \BibitemShut {NoStop}%
	\bibitem [{\citenamefont {{Huhtinen}}\ \emph {et~al.}(2022)\citenamefont
		{{Huhtinen}}, \citenamefont {{Herzog-Arbeitman}}, \citenamefont {{Chew}},
		\citenamefont {{Bernevig}},\ and\ \citenamefont {{T{\"o}rm{\"a}}}}]{FB-SC_2}%
	\BibitemOpen
	\bibfield  {author} {\bibinfo {author} {\bibfnamefont {Kukka-Emilia}\
			\bibnamefont {{Huhtinen}}}, \bibinfo {author} {\bibfnamefont {Jonah}\
			\bibnamefont {{Herzog-Arbeitman}}}, \bibinfo {author} {\bibfnamefont {Aaron}\
			\bibnamefont {{Chew}}}, \bibinfo {author} {\bibfnamefont {Bogdan~A.}\
			\bibnamefont {{Bernevig}}}, \ and\ \bibinfo {author} {\bibfnamefont
			{P{\"a}ivi}\ \bibnamefont {{T{\"o}rm{\"a}}}},\ }\bibfield  {title} {\enquote
		{\bibinfo {title} {{Revisiting flat band superconductivity: Dependence on
					minimal quantum metric and band touchings}},}\ }\href {\doibase
		10.1103/PhysRevB.106.014518} {\bibfield  {journal} {\bibinfo  {journal}
			{\prb}\ }\textbf {\bibinfo {volume} {106}},\ \bibinfo {eid} {014518}
		(\bibinfo {year} {2022})},\ \Eprint {http://arxiv.org/abs/2203.11133}
	{arXiv:2203.11133 [cond-mat.supr-con]} \BibitemShut {NoStop}%
	\bibitem [{\citenamefont {{Julku}}\ \emph {et~al.}(2020)\citenamefont
		{{Julku}}, \citenamefont {{Peltonen}}, \citenamefont {{Liang}}, \citenamefont
		{{Heikkil{\"a}}},\ and\ \citenamefont {{T{\"o}rm{\"a}}}}]{FB-SC_3}%
	\BibitemOpen
	\bibfield  {author} {\bibinfo {author} {\bibfnamefont {A.}~\bibnamefont
			{{Julku}}}, \bibinfo {author} {\bibfnamefont {T.~J.}\ \bibnamefont
			{{Peltonen}}}, \bibinfo {author} {\bibfnamefont {L.}~\bibnamefont {{Liang}}},
		\bibinfo {author} {\bibfnamefont {T.~T.}\ \bibnamefont {{Heikkil{\"a}}}}, \
		and\ \bibinfo {author} {\bibfnamefont {P.}~\bibnamefont {{T{\"o}rm{\"a}}}},\
	}\bibfield  {title} {\enquote {\bibinfo {title} {{Superfluid weight and
					Berezinskii-Kosterlitz-Thouless transition temperature of twisted bilayer
					graphene}},}\ }\href {\doibase 10.1103/PhysRevB.101.060505} {\bibfield
		{journal} {\bibinfo  {journal} {\prb}\ }\textbf {\bibinfo {volume} {101}},\
		\bibinfo {eid} {060505} (\bibinfo {year} {2020})},\ \Eprint
	{http://arxiv.org/abs/1906.06313} {arXiv:1906.06313 [cond-mat.mes-hall]}
	\BibitemShut {NoStop}%
	\bibitem [{\citenamefont {{Liang}}\ \emph {et~al.}(2017)\citenamefont
		{{Liang}}, \citenamefont {{Vanhala}}, \citenamefont {{Peotta}}, \citenamefont
		{{Siro}}, \citenamefont {{Harju}},\ and\ \citenamefont
		{{T{\"o}rm{\"a}}}}]{FB-SC_4}%
	\BibitemOpen
	\bibfield  {author} {\bibinfo {author} {\bibfnamefont {Long}\ \bibnamefont
			{{Liang}}}, \bibinfo {author} {\bibfnamefont {Tuomas~I.}\ \bibnamefont
			{{Vanhala}}}, \bibinfo {author} {\bibfnamefont {Sebastiano}\ \bibnamefont
			{{Peotta}}}, \bibinfo {author} {\bibfnamefont {Topi}\ \bibnamefont {{Siro}}},
		\bibinfo {author} {\bibfnamefont {Ari}\ \bibnamefont {{Harju}}}, \ and\
		\bibinfo {author} {\bibfnamefont {P{\"a}ivi}\ \bibnamefont
			{{T{\"o}rm{\"a}}}},\ }\bibfield  {title} {\enquote {\bibinfo {title} {{Band
					geometry, Berry curvature, and superfluid weight}},}\ }\href {\doibase
		10.1103/PhysRevB.95.024515} {\bibfield  {journal} {\bibinfo  {journal}
			{\prb}\ }\textbf {\bibinfo {volume} {95}},\ \bibinfo {eid} {024515} (\bibinfo
		{year} {2017})},\ \Eprint {http://arxiv.org/abs/1610.01803} {arXiv:1610.01803
		[cond-mat.supr-con]} \BibitemShut {NoStop}%
	\bibitem [{\citenamefont {{T{\"o}rm{\"a}}}\ \emph {et~al.}(2018)\citenamefont
		{{T{\"o}rm{\"a}}}, \citenamefont {{Liang}},\ and\ \citenamefont
		{{Peotta}}}]{FB-SC_5}%
	\BibitemOpen
	\bibfield  {author} {\bibinfo {author} {\bibfnamefont {P.}~\bibnamefont
			{{T{\"o}rm{\"a}}}}, \bibinfo {author} {\bibfnamefont {L.}~\bibnamefont
			{{Liang}}}, \ and\ \bibinfo {author} {\bibfnamefont {S.}~\bibnamefont
			{{Peotta}}},\ }\bibfield  {title} {\enquote {\bibinfo {title} {{Quantum
					metric and effective mass of a two-body bound state in a flat band}},}\
	}\href {\doibase 10.1103/PhysRevB.98.220511} {\bibfield  {journal} {\bibinfo
			{journal} {\prb}\ }\textbf {\bibinfo {volume} {98}},\ \bibinfo {eid} {220511}
		(\bibinfo {year} {2018})},\ \Eprint {http://arxiv.org/abs/1810.09870}
	{arXiv:1810.09870 [cond-mat.supr-con]} \BibitemShut {NoStop}%
	\bibitem [{\citenamefont {{Peotta}}\ and\ \citenamefont
		{{T{\"o}rm{\"a}}}(2015)}]{FB_1}%
	\BibitemOpen
	\bibfield  {author} {\bibinfo {author} {\bibfnamefont {Sebastiano}\
			\bibnamefont {{Peotta}}}\ and\ \bibinfo {author} {\bibfnamefont {P{\"a}ivi}\
			\bibnamefont {{T{\"o}rm{\"a}}}},\ }\bibfield  {title} {\enquote {\bibinfo
			{title} {{Superfluidity in topologically nontrivial flat bands}},}\ }\href
	{\doibase 10.1038/ncomms9944} {\bibfield  {journal} {\bibinfo  {journal}
			{Nature Communications}\ }\textbf {\bibinfo {volume} {6}},\ \bibinfo {eid}
		{8944} (\bibinfo {year} {2015})},\ \Eprint {http://arxiv.org/abs/1506.02815}
	{arXiv:1506.02815 [cond-mat.supr-con]} \BibitemShut {NoStop}%
	\bibitem [{\citenamefont {{T{\"o}rm{\"a}}}\ \emph {et~al.}(2022)\citenamefont
		{{T{\"o}rm{\"a}}}, \citenamefont {{Peotta}},\ and\ \citenamefont
		{{Bernevig}}}]{FB_2}%
	\BibitemOpen
	\bibfield  {author} {\bibinfo {author} {\bibfnamefont {P{\"a}ivi}\
			\bibnamefont {{T{\"o}rm{\"a}}}}, \bibinfo {author} {\bibfnamefont
			{Sebastiano}\ \bibnamefont {{Peotta}}}, \ and\ \bibinfo {author}
		{\bibfnamefont {Bogdan~A.}\ \bibnamefont {{Bernevig}}},\ }\bibfield  {title}
	{\enquote {\bibinfo {title} {{Superconductivity, superfluidity and quantum
					geometry in twisted multilayer systems}},}\ }\href {\doibase
		10.1038/s42254-022-00466-y} {\bibfield  {journal} {\bibinfo  {journal}
			{Nature Reviews Physics}\ }\textbf {\bibinfo {volume} {4}},\ \bibinfo {pages}
		{528--542} (\bibinfo {year} {2022})},\ \Eprint
	{http://arxiv.org/abs/2111.00807} {arXiv:2111.00807 [cond-mat.supr-con]}
	\BibitemShut {NoStop}%
	\bibitem [{\citenamefont {{Balents}}\ \emph {et~al.}(2020)\citenamefont
		{{Balents}}, \citenamefont {{Dean}}, \citenamefont {{Efetov}},\ and\
		\citenamefont {{Young}}}]{FB-SC-6}%
	\BibitemOpen
	\bibfield  {author} {\bibinfo {author} {\bibfnamefont {Leon}\ \bibnamefont
			{{Balents}}}, \bibinfo {author} {\bibfnamefont {Cory~R.}\ \bibnamefont
			{{Dean}}}, \bibinfo {author} {\bibfnamefont {Dmitri~K.}\ \bibnamefont
			{{Efetov}}}, \ and\ \bibinfo {author} {\bibfnamefont {Andrea~F.}\
			\bibnamefont {{Young}}},\ }\bibfield  {title} {\enquote {\bibinfo {title}
			{{Superconductivity and strong correlations in moir{\'e} flat bands}},}\
	}\href {\doibase 10.1038/s41567-020-0906-9} {\bibfield  {journal} {\bibinfo
			{journal} {Nature Physics}\ }\textbf {\bibinfo {volume} {16}},\ \bibinfo
		{pages} {725--733} (\bibinfo {year} {2020})}\BibitemShut {NoStop}%
	\bibitem [{\citenamefont {{Peri}}\ \emph {et~al.}(2021)\citenamefont {{Peri}},
		\citenamefont {{Song}}, \citenamefont {{Bernevig}},\ and\ \citenamefont
		{{Huber}}}]{FB-SC-7}%
	\BibitemOpen
	\bibfield  {author} {\bibinfo {author} {\bibfnamefont {Valerio}\ \bibnamefont
			{{Peri}}}, \bibinfo {author} {\bibfnamefont {Zhi-Da}\ \bibnamefont {{Song}}},
		\bibinfo {author} {\bibfnamefont {B.~Andrei}\ \bibnamefont {{Bernevig}}}, \
		and\ \bibinfo {author} {\bibfnamefont {Sebastian~D.}\ \bibnamefont
			{{Huber}}},\ }\bibfield  {title} {\enquote {\bibinfo {title} {{Fragile
					Topology and Flat-Band Superconductivity in the Strong-Coupling Regime}},}\
	}\href {\doibase 10.1103/PhysRevLett.126.027002} {\bibfield  {journal}
		{\bibinfo  {journal} {\prl}\ }\textbf {\bibinfo {volume} {126}},\ \bibinfo
		{eid} {027002} (\bibinfo {year} {2021})},\ \Eprint
	{http://arxiv.org/abs/2008.02288} {arXiv:2008.02288 [cond-mat.supr-con]}
	\BibitemShut {NoStop}%
	\bibitem [{\citenamefont {{Tang}}\ \emph {et~al.}(2020)\citenamefont {{Tang}},
		\citenamefont {{Li}}, \citenamefont {{Li}}, \citenamefont {{Xu}},
		\citenamefont {{Liu}}, \citenamefont {{Barmak}}, \citenamefont {{Watanabe}},
		\citenamefont {{Taniguchi}}, \citenamefont {{MacDonald}}, \citenamefont
		{{Shan}},\ and\ \citenamefont {{Mak}}}]{FM_1}%
	\BibitemOpen
	\bibfield  {author} {\bibinfo {author} {\bibfnamefont {Yanhao}\ \bibnamefont
			{{Tang}}}, \bibinfo {author} {\bibfnamefont {Lizhong}\ \bibnamefont {{Li}}},
		\bibinfo {author} {\bibfnamefont {Tingxin}\ \bibnamefont {{Li}}}, \bibinfo
		{author} {\bibfnamefont {Yang}\ \bibnamefont {{Xu}}}, \bibinfo {author}
		{\bibfnamefont {Song}\ \bibnamefont {{Liu}}}, \bibinfo {author}
		{\bibfnamefont {Katayun}\ \bibnamefont {{Barmak}}}, \bibinfo {author}
		{\bibfnamefont {Kenji}\ \bibnamefont {{Watanabe}}}, \bibinfo {author}
		{\bibfnamefont {Takashi}\ \bibnamefont {{Taniguchi}}}, \bibinfo {author}
		{\bibfnamefont {Allan~H.}\ \bibnamefont {{MacDonald}}}, \bibinfo {author}
		{\bibfnamefont {Jie}\ \bibnamefont {{Shan}}}, \ and\ \bibinfo {author}
		{\bibfnamefont {Kin~Fai}\ \bibnamefont {{Mak}}},\ }\bibfield  {title}
	{\enquote {\bibinfo {title} {{Simulation of Hubbard model physics in
					WSe$_{2}$/WS$_{2}$ moir{\'e} superlattices}},}\ }\href {\doibase
		10.1038/s41586-020-2085-3} {\bibfield  {journal} {\bibinfo  {journal} {\nat}\
		}\textbf {\bibinfo {volume} {579}},\ \bibinfo {pages} {353--358} (\bibinfo
		{year} {2020})}\BibitemShut {NoStop}%
	\bibitem [{\citenamefont {{Chen}}\ \emph {et~al.}(2020)\citenamefont {{Chen}},
		\citenamefont {{Sharpe}}, \citenamefont {{Fox}}, \citenamefont {{Zhang}},
		\citenamefont {{Wang}}, \citenamefont {{Jiang}}, \citenamefont {{Lyu}},
		\citenamefont {{Li}}, \citenamefont {{Watanabe}}, \citenamefont
		{{Taniguchi}}, \citenamefont {{Shi}}, \citenamefont {{Senthil}},
		\citenamefont {{Goldhaber-Gordon}}, \citenamefont {{Zhang}},\ and\
		\citenamefont {{Wang}}}]{FM_2}%
	\BibitemOpen
	\bibfield  {author} {\bibinfo {author} {\bibfnamefont {Guorui}\ \bibnamefont
			{{Chen}}}, \bibinfo {author} {\bibfnamefont {Aaron~L.}\ \bibnamefont
			{{Sharpe}}}, \bibinfo {author} {\bibfnamefont {Eli~J.}\ \bibnamefont
			{{Fox}}}, \bibinfo {author} {\bibfnamefont {Ya-Hui}\ \bibnamefont {{Zhang}}},
		\bibinfo {author} {\bibfnamefont {Shaoxin}\ \bibnamefont {{Wang}}}, \bibinfo
		{author} {\bibfnamefont {Lili}\ \bibnamefont {{Jiang}}}, \bibinfo {author}
		{\bibfnamefont {Bosai}\ \bibnamefont {{Lyu}}}, \bibinfo {author}
		{\bibfnamefont {Hongyuan}\ \bibnamefont {{Li}}}, \bibinfo {author}
		{\bibfnamefont {Kenji}\ \bibnamefont {{Watanabe}}}, \bibinfo {author}
		{\bibfnamefont {Takashi}\ \bibnamefont {{Taniguchi}}}, \bibinfo {author}
		{\bibfnamefont {Zhiwen}\ \bibnamefont {{Shi}}}, \bibinfo {author}
		{\bibfnamefont {T.}~\bibnamefont {{Senthil}}}, \bibinfo {author}
		{\bibfnamefont {David}\ \bibnamefont {{Goldhaber-Gordon}}}, \bibinfo {author}
		{\bibfnamefont {Yuanbo}\ \bibnamefont {{Zhang}}}, \ and\ \bibinfo {author}
		{\bibfnamefont {Feng}\ \bibnamefont {{Wang}}},\ }\bibfield  {title} {\enquote
		{\bibinfo {title} {{Tunable correlated Chern insulator and ferromagnetism in
					a moir{\'e} superlattice}},}\ }\href {\doibase 10.1038/s41586-020-2049-7}
	{\bibfield  {journal} {\bibinfo  {journal} {\nat}\ }\textbf {\bibinfo
			{volume} {579}},\ \bibinfo {pages} {56--61} (\bibinfo {year} {2020})},\
	\Eprint {http://arxiv.org/abs/1905.06535} {arXiv:1905.06535
		[cond-mat.mes-hall]} \BibitemShut {NoStop}%
	\bibitem [{\citenamefont {{Wu}}\ \emph {et~al.}(2025)\citenamefont {{Wu}},
		\citenamefont {{Xu}}, \citenamefont {{Wang}}, \citenamefont {{Lin}},
		\citenamefont {{Cao}},\ and\ \citenamefont {{Cao}}}]{FB-AFM-2}%
	\BibitemOpen
	\bibfield  {author} {\bibinfo {author} {\bibfnamefont {Siqi}\ \bibnamefont
			{{Wu}}}, \bibinfo {author} {\bibfnamefont {Chenchao}\ \bibnamefont {{Xu}}},
		\bibinfo {author} {\bibfnamefont {Xiaoqun}\ \bibnamefont {{Wang}}}, \bibinfo
		{author} {\bibfnamefont {Hai-Qing}\ \bibnamefont {{Lin}}}, \bibinfo {author}
		{\bibfnamefont {Chao}\ \bibnamefont {{Cao}}}, \ and\ \bibinfo {author}
		{\bibfnamefont {Guang-Han}\ \bibnamefont {{Cao}}},\ }\bibfield  {title}
	{\enquote {\bibinfo {title} {{Flat-band enhanced antiferromagnetic
					fluctuations and superconductivity in pressurized CsCr$_{3}$Sb$_{5}$}},}\
	}\href {\doibase 10.1038/s41467-025-56582-7} {\bibfield  {journal} {\bibinfo
			{journal} {Nature Communications}\ }\textbf {\bibinfo {volume} {16}},\
		\bibinfo {eid} {1375} (\bibinfo {year} {2025})},\ \Eprint
	{http://arxiv.org/abs/2404.04701} {arXiv:2404.04701 [cond-mat.supr-con]}
	\BibitemShut {NoStop}%
	\bibitem [{\citenamefont {{Alexeev}}\ \emph {et~al.}(2019)\citenamefont
		{{Alexeev}}, \citenamefont {{Ruiz-Tijerina}}, \citenamefont {{Danovich}},
		\citenamefont {{Hamer}}, \citenamefont {{Terry}}, \citenamefont {{Nayak}},
		\citenamefont {{Ahn}}, \citenamefont {{Pak}}, \citenamefont {{Lee}},
		\citenamefont {{Sohn}}, \citenamefont {{Molas}}, \citenamefont {{Koperski}},
		\citenamefont {{Watanabe}}, \citenamefont {{Taniguchi}}, \citenamefont
		{{Novoselov}}, \citenamefont {{Gorbachev}}, \citenamefont {{Shin}},
		\citenamefont {{Fal'ko}},\ and\ \citenamefont {{Tartakovskii}}}]{Ex_1}%
	\BibitemOpen
	\bibfield  {author} {\bibinfo {author} {\bibfnamefont {Evgeny~M.}\
			\bibnamefont {{Alexeev}}}, \bibinfo {author} {\bibfnamefont {David~A.}\
			\bibnamefont {{Ruiz-Tijerina}}}, \bibinfo {author} {\bibfnamefont {Mark}\
			\bibnamefont {{Danovich}}}, \bibinfo {author} {\bibfnamefont {Matthew~J.}\
			\bibnamefont {{Hamer}}}, \bibinfo {author} {\bibfnamefont {Daniel~J.}\
			\bibnamefont {{Terry}}}, \bibinfo {author} {\bibfnamefont {Pramoda~K.}\
			\bibnamefont {{Nayak}}}, \bibinfo {author} {\bibfnamefont {Seongjoon}\
			\bibnamefont {{Ahn}}}, \bibinfo {author} {\bibfnamefont {Sangyeon}\
			\bibnamefont {{Pak}}}, \bibinfo {author} {\bibfnamefont {Juwon}\ \bibnamefont
			{{Lee}}}, \bibinfo {author} {\bibfnamefont {Jung~Inn}\ \bibnamefont
			{{Sohn}}}, \bibinfo {author} {\bibfnamefont {Maciej~R.}\ \bibnamefont
			{{Molas}}}, \bibinfo {author} {\bibfnamefont {Maciej}\ \bibnamefont
			{{Koperski}}}, \bibinfo {author} {\bibfnamefont {Kenji}\ \bibnamefont
			{{Watanabe}}}, \bibinfo {author} {\bibfnamefont {Takashi}\ \bibnamefont
			{{Taniguchi}}}, \bibinfo {author} {\bibfnamefont {Kostya~S.}\ \bibnamefont
			{{Novoselov}}}, \bibinfo {author} {\bibfnamefont {Roman~V.}\ \bibnamefont
			{{Gorbachev}}}, \bibinfo {author} {\bibfnamefont {Hyeon~Suk}\ \bibnamefont
			{{Shin}}}, \bibinfo {author} {\bibfnamefont {Vladimir~I.}\ \bibnamefont
			{{Fal'ko}}}, \ and\ \bibinfo {author} {\bibfnamefont {Alexander~I.}\
			\bibnamefont {{Tartakovskii}}},\ }\bibfield  {title} {\enquote {\bibinfo
			{title} {{Resonantly hybridized excitons in moir{\'e} superlattices in van
					der Waals heterostructures}},}\ }\href {\doibase 10.1038/s41586-019-0986-9}
	{\bibfield  {journal} {\bibinfo  {journal} {\nat}\ }\textbf {\bibinfo
			{volume} {567}},\ \bibinfo {pages} {81--86} (\bibinfo {year} {2019})},\
	\Eprint {http://arxiv.org/abs/1904.06214} {arXiv:1904.06214
		[cond-mat.mes-hall]} \BibitemShut {NoStop}%
	\bibitem [{\citenamefont {{Rivera}}\ \emph {et~al.}(2018)\citenamefont
		{{Rivera}}, \citenamefont {{Yu}}, \citenamefont {{Seyler}}, \citenamefont
		{{Wilson}}, \citenamefont {{Yao}},\ and\ \citenamefont {{Xu}}}]{Ex_2}%
	\BibitemOpen
	\bibfield  {author} {\bibinfo {author} {\bibfnamefont {Pasqual}\ \bibnamefont
			{{Rivera}}}, \bibinfo {author} {\bibfnamefont {Hongyi}\ \bibnamefont {{Yu}}},
		\bibinfo {author} {\bibfnamefont {Kyle~L.}\ \bibnamefont {{Seyler}}},
		\bibinfo {author} {\bibfnamefont {Nathan~P.}\ \bibnamefont {{Wilson}}},
		\bibinfo {author} {\bibfnamefont {Wang}\ \bibnamefont {{Yao}}}, \ and\
		\bibinfo {author} {\bibfnamefont {Xiaodong}\ \bibnamefont {{Xu}}},\
	}\bibfield  {title} {\enquote {\bibinfo {title} {{Interlayer valley excitons
					in heterobilayers of transition metal dichalcogenides}},}\ }\href {\doibase
		10.1038/s41565-018-0193-0} {\bibfield  {journal} {\bibinfo  {journal} {Nature
				Nanotechnology}\ }\textbf {\bibinfo {volume} {13}},\ \bibinfo {pages}
		{1004--1015} (\bibinfo {year} {2018})}\BibitemShut {NoStop}%
	\bibitem [{\citenamefont {{Ying}}\ and\ \citenamefont
		{{Law}}(2024)}]{FB-exciton}%
	\BibitemOpen
	\bibfield  {author} {\bibinfo {author} {\bibfnamefont {Xuzhe}\ \bibnamefont
			{{Ying}}}\ and\ \bibinfo {author} {\bibfnamefont {K.~T.}\ \bibnamefont
			{{Law}}},\ }\bibfield  {title} {\enquote {\bibinfo {title} {{Flat band
					excitons and quantum metric}},}\ }\href {\doibase 10.48550/arXiv.2407.00325}
	{\bibfield  {journal} {\bibinfo  {journal} {arXiv e-prints}\ ,\ \bibinfo
			{eid} {arXiv:2407.00325}} (\bibinfo {year} {2024})},\ \Eprint
	{http://arxiv.org/abs/2407.00325} {arXiv:2407.00325 [cond-mat.mes-hall]}
	\BibitemShut {NoStop}%
	\bibitem [{\citenamefont {{Provost}}\ and\ \citenamefont
		{{Vallee}}(1980)}]{QMT}%
	\BibitemOpen
	\bibfield  {author} {\bibinfo {author} {\bibfnamefont {J.~P.}\ \bibnamefont
			{{Provost}}}\ and\ \bibinfo {author} {\bibfnamefont {G.}~\bibnamefont
			{{Vallee}}},\ }\bibfield  {title} {\enquote {\bibinfo {title} {{Riemannian
					structure on manifolds of quantum states}},}\ }\href {\doibase
		10.1007/BF02193559} {\bibfield  {journal} {\bibinfo  {journal}
			{Communications in Mathematical Physics}\ }\textbf {\bibinfo {volume} {76}},\
		\bibinfo {pages} {289--301} (\bibinfo {year} {1980})}\BibitemShut {NoStop}%
	\bibitem [{\citenamefont {{Anandan}}\ and\ \citenamefont
		{{Aharonov}}(1990)}]{QM_D_1}%
	\BibitemOpen
	\bibfield  {author} {\bibinfo {author} {\bibfnamefont {J.}~\bibnamefont
			{{Anandan}}}\ and\ \bibinfo {author} {\bibfnamefont {Y.}~\bibnamefont
			{{Aharonov}}},\ }\bibfield  {title} {\enquote {\bibinfo {title} {{Geometry of
					quantum evolution}},}\ }\href {\doibase 10.1103/PhysRevLett.65.1697}
	{\bibfield  {journal} {\bibinfo  {journal} {\prl}\ }\textbf {\bibinfo
			{volume} {65}},\ \bibinfo {pages} {1697--1700} (\bibinfo {year}
		{1990})}\BibitemShut {NoStop}%
	\bibitem [{\citenamefont {{Resta}}(2011)}]{QM_D_2}%
	\BibitemOpen
	\bibfield  {author} {\bibinfo {author} {\bibfnamefont {R.}~\bibnamefont
			{{Resta}}},\ }\bibfield  {title} {\enquote {\bibinfo {title} {{The insulating
					state of matter: a geometrical theory}},}\ }\href {\doibase
		10.1140/epjb/e2010-10874-4} {\bibfield  {journal} {\bibinfo  {journal}
			{European Physical Journal B}\ }\textbf {\bibinfo {volume} {79}},\ \bibinfo
		{pages} {121--137} (\bibinfo {year} {2011})},\ \Eprint
	{http://arxiv.org/abs/1012.5776} {arXiv:1012.5776 [cond-mat.mtrl-sci]}
	\BibitemShut {NoStop}%
	\bibitem [{\citenamefont {{Liu}}\ \emph {et~al.}(2025)\citenamefont {{Liu}},
		\citenamefont {{Qiang}}, \citenamefont {{Lu}},\ and\ \citenamefont
		{{Xie}}}]{QM-Cond-Mat}%
	\BibitemOpen
	\bibfield  {author} {\bibinfo {author} {\bibfnamefont {Tianyu}\ \bibnamefont
			{{Liu}}}, \bibinfo {author} {\bibfnamefont {Xiao-Bin}\ \bibnamefont
			{{Qiang}}}, \bibinfo {author} {\bibfnamefont {Hai-Zhou}\ \bibnamefont
			{{Lu}}}, \ and\ \bibinfo {author} {\bibfnamefont {X.~C.}\ \bibnamefont
			{{Xie}}},\ }\bibfield  {title} {\enquote {\bibinfo {title} {{Quantum geometry
					in condensed matter}},}\ }\href {\doibase 10.1093/nsr/nwae334} {\bibfield
		{journal} {\bibinfo  {journal} {National Science Review}\ }\textbf {\bibinfo
			{volume} {12}},\ \bibinfo {eid} {nwae334} (\bibinfo {year} {2025})},\ \Eprint
	{http://arxiv.org/abs/2409.13408} {arXiv:2409.13408 [cond-mat.mes-hall]}
	\BibitemShut {NoStop}%
	\bibitem [{\citenamefont {{Breach}}\ \emph {et~al.}(2024)\citenamefont
		{{Breach}}, \citenamefont {{Slager}},\ and\ \citenamefont
		{{{\"U}nal}}}]{Berry_1}%
	\BibitemOpen
	\bibfield  {author} {\bibinfo {author} {\bibfnamefont {Oliver}\ \bibnamefont
			{{Breach}}}, \bibinfo {author} {\bibfnamefont {Robert-Jan}\ \bibnamefont
			{{Slager}}}, \ and\ \bibinfo {author} {\bibfnamefont {F.~Nur}\ \bibnamefont
			{{{\"U}nal}}},\ }\bibfield  {title} {\enquote {\bibinfo {title}
			{{Interferometry of Non-Abelian Band Singularities and Euler Class
					Topology}},}\ }\href {\doibase 10.1103/PhysRevLett.133.093404} {\bibfield
		{journal} {\bibinfo  {journal} {\prl}\ }\textbf {\bibinfo {volume} {133}},\
		\bibinfo {eid} {093404} (\bibinfo {year} {2024})},\ \Eprint
	{http://arxiv.org/abs/2401.01928} {arXiv:2401.01928 [cond-mat.quant-gas]}
	\BibitemShut {NoStop}%
	\bibitem [{\citenamefont {Jankowski}\ \emph {et~al.}(2025)\citenamefont
		{Jankowski}, \citenamefont {Morris}, \citenamefont {Bouhon}, \citenamefont
		{\"Unal},\ and\ \citenamefont {Slager}}]{Berry_2}%
	\BibitemOpen
	\bibfield  {author} {\bibinfo {author} {\bibfnamefont {Wojciech~J.}\
			\bibnamefont {Jankowski}}, \bibinfo {author} {\bibfnamefont {Arthur~S.}\
			\bibnamefont {Morris}}, \bibinfo {author} {\bibfnamefont {Adrien}\
			\bibnamefont {Bouhon}}, \bibinfo {author} {\bibfnamefont {F.~Nur}\
			\bibnamefont {\"Unal}}, \ and\ \bibinfo {author} {\bibfnamefont {Robert-Jan}\
			\bibnamefont {Slager}},\ }\bibfield  {title} {\enquote {\bibinfo {title}
			{Optical manifestations and bounds of topological euler class},}\ }\href
	{\doibase 10.1103/PhysRevB.111.L081103} {\bibfield  {journal} {\bibinfo
			{journal} {Phys. Rev. B}\ }\textbf {\bibinfo {volume} {111}},\ \bibinfo
		{pages} {L081103} (\bibinfo {year} {2025})}\BibitemShut {NoStop}%
	\bibitem [{\citenamefont {{Marzari}}\ \emph {et~al.}(2012)\citenamefont
		{{Marzari}}, \citenamefont {{Mostofi}}, \citenamefont {{Yates}},
		\citenamefont {{Souza}},\ and\ \citenamefont
		{{Vanderbilt}}}]{Max-Localize-1}%
	\BibitemOpen
	\bibfield  {author} {\bibinfo {author} {\bibfnamefont {Nicola}\ \bibnamefont
			{{Marzari}}}, \bibinfo {author} {\bibfnamefont {Arash~A.}\ \bibnamefont
			{{Mostofi}}}, \bibinfo {author} {\bibfnamefont {Jonathan~R.}\ \bibnamefont
			{{Yates}}}, \bibinfo {author} {\bibfnamefont {Ivo}\ \bibnamefont {{Souza}}},
		\ and\ \bibinfo {author} {\bibfnamefont {David}\ \bibnamefont
			{{Vanderbilt}}},\ }\bibfield  {title} {\enquote {\bibinfo {title} {{Maximally
					localized Wannier functions: Theory and applications}},}\ }\href {\doibase
		10.1103/RevModPhys.84.1419} {\bibfield  {journal} {\bibinfo  {journal}
			{Reviews of Modern Physics}\ }\textbf {\bibinfo {volume} {84}},\ \bibinfo
		{pages} {1419--1475} (\bibinfo {year} {2012})},\ \Eprint
	{http://arxiv.org/abs/1112.5411} {arXiv:1112.5411 [cond-mat.mtrl-sci]}
	\BibitemShut {NoStop}%
	\bibitem [{\citenamefont {{Marzari}}\ and\ \citenamefont
		{{Vanderbilt}}(1997)}]{Max-Localize-2}%
	\BibitemOpen
	\bibfield  {author} {\bibinfo {author} {\bibfnamefont {Nicola}\ \bibnamefont
			{{Marzari}}}\ and\ \bibinfo {author} {\bibfnamefont {David}\ \bibnamefont
			{{Vanderbilt}}},\ }\bibfield  {title} {\enquote {\bibinfo {title} {{Maximally
					localized generalized Wannier functions for composite energy bands}},}\
	}\href {\doibase 10.1103/PhysRevB.56.12847} {\bibfield  {journal} {\bibinfo
			{journal} {\prb}\ }\textbf {\bibinfo {volume} {56}},\ \bibinfo {pages}
		{12847--12865} (\bibinfo {year} {1997})},\ \Eprint
	{http://arxiv.org/abs/cond-mat/9707145} {arXiv:cond-mat/9707145
		[cond-mat.mtrl-sci]} \BibitemShut {NoStop}%
	\bibitem [{\citenamefont {{Chen}}\ and\ \citenamefont {{Law}}(2024)}]{QM_SC_1}%
	\BibitemOpen
	\bibfield  {author} {\bibinfo {author} {\bibfnamefont {Shuai~A.}\
			\bibnamefont {{Chen}}}\ and\ \bibinfo {author} {\bibfnamefont {K.~T.}\
			\bibnamefont {{Law}}},\ }\bibfield  {title} {\enquote {\bibinfo {title}
			{{Ginzburg-Landau Theory of Flat-Band Superconductors with Quantum
					Metric}},}\ }\href {\doibase 10.1103/PhysRevLett.132.026002} {\bibfield
		{journal} {\bibinfo  {journal} {\prl}\ }\textbf {\bibinfo {volume} {132}},\
		\bibinfo {eid} {026002} (\bibinfo {year} {2024})},\ \Eprint
	{http://arxiv.org/abs/2303.15504} {arXiv:2303.15504 [cond-mat.supr-con]}
	\BibitemShut {NoStop}%
	\bibitem [{\citenamefont {{Hu}}\ \emph {et~al.}(2023)\citenamefont {{Hu}},
		\citenamefont {{Chen}},\ and\ \citenamefont {{Law}}}]{QM_SC_2}%
	\BibitemOpen
	\bibfield  {author} {\bibinfo {author} {\bibfnamefont {Jin-Xin}\ \bibnamefont
			{{Hu}}}, \bibinfo {author} {\bibfnamefont {Shuai~A.}\ \bibnamefont {{Chen}}},
		\ and\ \bibinfo {author} {\bibfnamefont {K.~T.}\ \bibnamefont {{Law}}},\
	}\bibfield  {title} {\enquote {\bibinfo {title} {{Anomalous Coherence Length
					in Superconductors with Quantum Metric}},}\ }\href {\doibase
		10.48550/arXiv.2308.05686} {\bibfield  {journal} {\bibinfo  {journal} {arXiv
				e-prints}\ ,\ \bibinfo {eid} {arXiv:2308.05686}} (\bibinfo {year} {2023})},\
	\Eprint {http://arxiv.org/abs/2308.05686} {arXiv:2308.05686
		[cond-mat.supr-con]} \BibitemShut {NoStop}%
	\bibitem [{\citenamefont {{Antebi}}\ \emph {et~al.}(2024)\citenamefont
		{{Antebi}}, \citenamefont {{Mitscherling}},\ and\ \citenamefont
		{{Holder}}}]{Int_1}%
	\BibitemOpen
	\bibfield  {author} {\bibinfo {author} {\bibfnamefont {Ohad}\ \bibnamefont
			{{Antebi}}}, \bibinfo {author} {\bibfnamefont {Johannes}\ \bibnamefont
			{{Mitscherling}}}, \ and\ \bibinfo {author} {\bibfnamefont {Tobias}\
			\bibnamefont {{Holder}}},\ }\bibfield  {title} {\enquote {\bibinfo {title}
			{{Drude weight of an interacting flat-band metal}},}\ }\href {\doibase
		10.1103/PhysRevB.110.L241111} {\bibfield  {journal} {\bibinfo  {journal}
			{\prb}\ }\textbf {\bibinfo {volume} {110}},\ \bibinfo {eid} {L241111}
		(\bibinfo {year} {2024})},\ \Eprint {http://arxiv.org/abs/2407.09599}
	{arXiv:2407.09599 [cond-mat.str-el]} \BibitemShut {NoStop}%
	\bibitem [{\citenamefont {{Mitscherling}}\ and\ \citenamefont
		{{Holder}}(2022)}]{Dis_1}%
	\BibitemOpen
	\bibfield  {author} {\bibinfo {author} {\bibfnamefont {Johannes}\
			\bibnamefont {{Mitscherling}}}\ and\ \bibinfo {author} {\bibfnamefont
			{Tobias}\ \bibnamefont {{Holder}}},\ }\bibfield  {title} {\enquote {\bibinfo
			{title} {{Bound on resistivity in flat-band materials due to the quantum
					metric}},}\ }\href {\doibase 10.1103/PhysRevB.105.085154} {\bibfield
		{journal} {\bibinfo  {journal} {\prb}\ }\textbf {\bibinfo {volume} {105}},\
		\bibinfo {eid} {085154} (\bibinfo {year} {2022})},\ \Eprint
	{http://arxiv.org/abs/2110.14658} {arXiv:2110.14658 [cond-mat.mes-hall]}
	\BibitemShut {NoStop}%
	\bibitem [{\citenamefont {{Mera}}\ and\ \citenamefont
		{{Mitscherling}}(2022)}]{FB-deg-1}%
	\BibitemOpen
	\bibfield  {author} {\bibinfo {author} {\bibfnamefont {Bruno}\ \bibnamefont
			{{Mera}}}\ and\ \bibinfo {author} {\bibfnamefont {Johannes}\ \bibnamefont
			{{Mitscherling}}},\ }\bibfield  {title} {\enquote {\bibinfo {title}
			{{Nontrivial quantum geometry of degenerate flat bands}},}\ }\href {\doibase
		10.1103/PhysRevB.106.165133} {\bibfield  {journal} {\bibinfo  {journal}
			{\prb}\ }\textbf {\bibinfo {volume} {106}},\ \bibinfo {eid} {165133}
		(\bibinfo {year} {2022})},\ \Eprint {http://arxiv.org/abs/2205.07900}
	{arXiv:2205.07900 [cond-mat.mes-hall]} \BibitemShut {NoStop}%
	\bibitem [{\citenamefont {{Bouzerar}}(2022)}]{QM-FB-Dis_1}%
	\BibitemOpen
	\bibfield  {author} {\bibinfo {author} {\bibfnamefont {G.}~\bibnamefont
			{{Bouzerar}}},\ }\bibfield  {title} {\enquote {\bibinfo {title} {{Giant boost
					of the quantum metric in disordered one-dimensional flat-band systems}},}\
	}\href {\doibase 10.1103/PhysRevB.106.125125} {\bibfield  {journal} {\bibinfo
			{journal} {\prb}\ }\textbf {\bibinfo {volume} {106}},\ \bibinfo {eid}
		{125125} (\bibinfo {year} {2022})},\ \Eprint
	{http://arxiv.org/abs/2205.06164} {arXiv:2205.06164 [quant-ph]} \BibitemShut
	{NoStop}%
	\bibitem [{\citenamefont {{Chen}}\ \emph {et~al.}(2025)\citenamefont {{Chen}},
		\citenamefont {{Moessner}},\ and\ \citenamefont {{Ng}}}]{xkxw-1134}%
	\BibitemOpen
	\bibfield  {author} {\bibinfo {author} {\bibfnamefont {Shuai~A.}\
			\bibnamefont {{Chen}}}, \bibinfo {author} {\bibfnamefont {Roderich}\
			\bibnamefont {{Moessner}}}, \ and\ \bibinfo {author} {\bibfnamefont
			{Tai~Kai}\ \bibnamefont {{Ng}}},\ }\bibfield  {title} {\enquote {\bibinfo
			{title} {{Generalized Peierls Substitution for Wannier Obstructions: Response
					to Disorder and Interactions}},}\ }\href {\doibase 10.1103/xkxw-1134}
	{\bibfield  {journal} {\bibinfo  {journal} {\prl}\ }\textbf {\bibinfo
			{volume} {135}},\ \bibinfo {eid} {116502} (\bibinfo {year} {2025})},\ \Eprint
	{http://arxiv.org/abs/2503.09709} {arXiv:2503.09709 [cond-mat.str-el]}
	\BibitemShut {NoStop}%
	\bibitem [{\citenamefont {{Rhim}}\ and\ \citenamefont
		{{Yang}}(2021)}]{FB-transport-2}%
	\BibitemOpen
	\bibfield  {author} {\bibinfo {author} {\bibfnamefont {Jun-Won}\ \bibnamefont
			{{Rhim}}}\ and\ \bibinfo {author} {\bibfnamefont {Bohm-Jung}\ \bibnamefont
			{{Yang}}},\ }\bibfield  {title} {\enquote {\bibinfo {title} {{Singular flat
					bands}},}\ }\href {\doibase 10.1080/23746149.2021.1901606} {\bibfield
		{journal} {\bibinfo  {journal} {Advances in Physics X}\ }\textbf {\bibinfo
			{volume} {6}},\ \bibinfo {pages} {1901606} (\bibinfo {year} {2021})},\
	\Eprint {http://arxiv.org/abs/2012.04279} {arXiv:2012.04279 [physics.optics]}
	\BibitemShut {NoStop}%
	\bibitem [{\citenamefont {{Onishi}}\ and\ \citenamefont
		{{Fu}}(2025)}]{QuantumWeight-1}%
	\BibitemOpen
	\bibfield  {author} {\bibinfo {author} {\bibfnamefont {Yugo}\ \bibnamefont
			{{Onishi}}}\ and\ \bibinfo {author} {\bibfnamefont {Liang}\ \bibnamefont
			{{Fu}}},\ }\bibfield  {title} {\enquote {\bibinfo {title} {{Quantum weight: A
					fundamental property of quantum many-body systems}},}\ }\href {\doibase
		10.1103/PhysRevResearch.7.023158} {\bibfield  {journal} {\bibinfo  {journal}
			{Physical Review Research}\ }\textbf {\bibinfo {volume} {7}},\ \bibinfo {eid}
		{023158} (\bibinfo {year} {2025})},\ \Eprint
	{http://arxiv.org/abs/2406.06783} {arXiv:2406.06783 [cond-mat.str-el]}
	\BibitemShut {NoStop}%
	\bibitem [{\citenamefont {{Kruchkov}}(2023)}]{FB-Trans-1}%
	\BibitemOpen
	\bibfield  {author} {\bibinfo {author} {\bibfnamefont {Alexander}\
			\bibnamefont {{Kruchkov}}},\ }\bibfield  {title} {\enquote {\bibinfo {title}
			{{Quantum transport anomalies in dispersionless quantum states}},}\ }\href
	{\doibase 10.1103/PhysRevB.107.L241102} {\bibfield  {journal} {\bibinfo
			{journal} {\prb}\ }\textbf {\bibinfo {volume} {107}},\ \bibinfo {eid}
		{L241102} (\bibinfo {year} {2023})}\BibitemShut {NoStop}%
	\bibitem [{\citenamefont {{Girvin}}\ and\ \citenamefont
		{{Yang}}(2019)}]{MCMP-Girvin}%
	\BibitemOpen
	\bibfield  {author} {\bibinfo {author} {\bibfnamefont {Steven~M.}\
			\bibnamefont {{Girvin}}}\ and\ \bibinfo {author} {\bibfnamefont {Kun}\
			\bibnamefont {{Yang}}},\ }\href {\doibase 10.1017/9781316480649} {\emph
		{\bibinfo {title} {{Modern Condensed Matter Physics}}}}\ (\bibinfo
	{publisher} {Cambridge University Press},\ \bibinfo {year}
	{2019})\BibitemShut {NoStop}%
	\bibitem [{\citenamefont {{Huhtinen}}\ and\ \citenamefont
		{{T{\"o}rm{\"a}}}(2023)}]{Kubo-Greenwood-1}%
	\BibitemOpen
	\bibfield  {author} {\bibinfo {author} {\bibfnamefont {Kukka-Emilia}\
			\bibnamefont {{Huhtinen}}}\ and\ \bibinfo {author} {\bibfnamefont
			{P{\"a}ivi}\ \bibnamefont {{T{\"o}rm{\"a}}}},\ }\bibfield  {title} {\enquote
		{\bibinfo {title} {{Conductivity in flat bands from the Kubo-Greenwood
					formula}},}\ }\href {\doibase 10.1103/PhysRevB.108.155108} {\bibfield
		{journal} {\bibinfo  {journal} {\prb}\ }\textbf {\bibinfo {volume} {108}},\
		\bibinfo {eid} {155108} (\bibinfo {year} {2023})},\ \Eprint
	{http://arxiv.org/abs/2212.03192} {arXiv:2212.03192 [cond-mat.mes-hall]}
	\BibitemShut {NoStop}%
	\bibitem [{\citenamefont {{Mitscherling}}(2020)}]{Dis_2}%
	\BibitemOpen
	\bibfield  {author} {\bibinfo {author} {\bibfnamefont {Johannes}\
			\bibnamefont {{Mitscherling}}},\ }\bibfield  {title} {\enquote {\bibinfo
			{title} {{Longitudinal and anomalous Hall conductivity of a general two-band
					model}},}\ }\href {\doibase 10.1103/PhysRevB.102.165151} {\bibfield
		{journal} {\bibinfo  {journal} {\prb}\ }\textbf {\bibinfo {volume} {102}},\
		\bibinfo {eid} {165151} (\bibinfo {year} {2020})},\ \Eprint
	{http://arxiv.org/abs/2008.11218} {arXiv:2008.11218 [cond-mat.str-el]}
	\BibitemShut {NoStop}%
	\bibitem [{\citenamefont {{Bouzerar}}\ and\ \citenamefont
		{{Mayou}}(2020)}]{FB-Nonzero-trans-1}%
	\BibitemOpen
	\bibfield  {author} {\bibinfo {author} {\bibfnamefont {G.}~\bibnamefont
			{{Bouzerar}}}\ and\ \bibinfo {author} {\bibfnamefont {D.}~\bibnamefont
			{{Mayou}}},\ }\bibfield  {title} {\enquote {\bibinfo {title} {{Quantum
					transport in self-similar graphene carpets}},}\ }\href {\doibase
		10.1103/PhysRevResearch.2.033063} {\bibfield  {journal} {\bibinfo  {journal}
			{Physical Review Research}\ }\textbf {\bibinfo {volume} {2}},\ \bibinfo {eid}
		{033063} (\bibinfo {year} {2020})}\BibitemShut {NoStop}%
	\bibitem [{\citenamefont {{Bouzerar}}\ and\ \citenamefont
		{{Mayou}}(2021)}]{QM-FB-Dis_2}%
	\BibitemOpen
	\bibfield  {author} {\bibinfo {author} {\bibfnamefont {G.}~\bibnamefont
			{{Bouzerar}}}\ and\ \bibinfo {author} {\bibfnamefont {D.}~\bibnamefont
			{{Mayou}}},\ }\bibfield  {title} {\enquote {\bibinfo {title} {{Quantum
					transport in flat bands and supermetallicity}},}\ }\href {\doibase
		10.1103/PhysRevB.103.075415} {\bibfield  {journal} {\bibinfo  {journal}
			{\prb}\ }\textbf {\bibinfo {volume} {103}},\ \bibinfo {eid} {075415}
		(\bibinfo {year} {2021})},\ \Eprint {http://arxiv.org/abs/2007.05309}
	{arXiv:2007.05309 [cond-mat.mes-hall]} \BibitemShut {NoStop}%
	\bibitem [{\citenamefont {{Laubscher}}\ \emph {et~al.}(2023)\citenamefont
		{{Laubscher}}, \citenamefont {{Weber}}, \citenamefont {{H{\"u}nenberger}},
		\citenamefont {{Schoeller}}, \citenamefont {{Kennes}}, \citenamefont
		{{Loss}},\ and\ \citenamefont {{Klinovaja}}}]{Int_2}%
	\BibitemOpen
	\bibfield  {author} {\bibinfo {author} {\bibfnamefont {Katharina}\
			\bibnamefont {{Laubscher}}}, \bibinfo {author} {\bibfnamefont {Clara~S.}\
			\bibnamefont {{Weber}}}, \bibinfo {author} {\bibfnamefont {Maximilian}\
			\bibnamefont {{H{\"u}nenberger}}}, \bibinfo {author} {\bibfnamefont
			{Herbert}\ \bibnamefont {{Schoeller}}}, \bibinfo {author} {\bibfnamefont
			{Dante~M.}\ \bibnamefont {{Kennes}}}, \bibinfo {author} {\bibfnamefont
			{Daniel}\ \bibnamefont {{Loss}}}, \ and\ \bibinfo {author} {\bibfnamefont
			{Jelena}\ \bibnamefont {{Klinovaja}}},\ }\bibfield  {title} {\enquote
		{\bibinfo {title} {{RKKY interaction in one-dimensional flat-band
					lattices}},}\ }\href {\doibase 10.1103/PhysRevB.108.155429} {\bibfield
		{journal} {\bibinfo  {journal} {\prb}\ }\textbf {\bibinfo {volume} {108}},\
		\bibinfo {eid} {155429} (\bibinfo {year} {2023})},\ \Eprint
	{http://arxiv.org/abs/2210.10025} {arXiv:2210.10025 [cond-mat.mes-hall]}
	\BibitemShut {NoStop}%
	\bibitem [{\citenamefont {{Checkelsky}}\ \emph {et~al.}(2023)\citenamefont
		{{Checkelsky}}, \citenamefont {{Bernevig}}, \citenamefont {{Coleman}},
		\citenamefont {{Si}},\ and\ \citenamefont {{Paschen}}}]{Int_3}%
	\BibitemOpen
	\bibfield  {author} {\bibinfo {author} {\bibfnamefont {Joseph~G.}\
			\bibnamefont {{Checkelsky}}}, \bibinfo {author} {\bibfnamefont {B.~Andrei}\
			\bibnamefont {{Bernevig}}}, \bibinfo {author} {\bibfnamefont {Piers}\
			\bibnamefont {{Coleman}}}, \bibinfo {author} {\bibfnamefont {Qimiao}\
			\bibnamefont {{Si}}}, \ and\ \bibinfo {author} {\bibfnamefont {Silke}\
			\bibnamefont {{Paschen}}},\ }\bibfield  {title} {\enquote {\bibinfo {title}
			{{Flat bands, strange metals, and the Kondo effect}},}\ }\href {\doibase
		10.48550/arXiv.2312.10659} {\bibfield  {journal} {\bibinfo  {journal} {arXiv
				e-prints}\ ,\ \bibinfo {eid} {arXiv:2312.10659}} (\bibinfo {year} {2023})},\
	\Eprint {http://arxiv.org/abs/2312.10659} {arXiv:2312.10659
		[cond-mat.str-el]} \BibitemShut {NoStop}%
	\bibitem [{\citenamefont {{Martinez}}\ \emph {et~al.}(2023)\citenamefont
		{{Martinez}}, \citenamefont {{Chiu}}, \citenamefont {{Smitham}},\ and\
		\citenamefont {{Houck}}}]{FB-Localization-1}%
	\BibitemOpen
	\bibfield  {author} {\bibinfo {author} {\bibfnamefont {Jeronimo G.~C.}\
			\bibnamefont {{Martinez}}}, \bibinfo {author} {\bibfnamefont {Christie~S.}\
			\bibnamefont {{Chiu}}}, \bibinfo {author} {\bibfnamefont {Basil~M.}\
			\bibnamefont {{Smitham}}}, \ and\ \bibinfo {author} {\bibfnamefont
			{Andrew~A.}\ \bibnamefont {{Houck}}},\ }\bibfield  {title} {\enquote
		{\bibinfo {title} {{Flat-band localization and interaction-induced
					delocalization of photons}},}\ }\href {\doibase 10.1126/sciadv.adj7195}
	{\bibfield  {journal} {\bibinfo  {journal} {Science Advances}\ }\textbf
		{\bibinfo {volume} {9}},\ \bibinfo {eid} {eadj7195} (\bibinfo {year}
		{2023})},\ \Eprint {http://arxiv.org/abs/2303.02170} {arXiv:2303.02170
		[quant-ph]} \BibitemShut {NoStop}%
	\bibitem [{\citenamefont {{Vidal}}\ \emph {et~al.}(2000)\citenamefont
		{{Vidal}}, \citenamefont {{Dou{\c{c}}ot}}, \citenamefont {{Mosseri}},\ and\
		\citenamefont {{Butaud}}}]{FB-Interatcion-1}%
	\BibitemOpen
	\bibfield  {author} {\bibinfo {author} {\bibfnamefont {Julien}\ \bibnamefont
			{{Vidal}}}, \bibinfo {author} {\bibfnamefont {Beno{\^\i}t}\ \bibnamefont
			{{Dou{\c{c}}ot}}}, \bibinfo {author} {\bibfnamefont {R{\'e}my}\ \bibnamefont
			{{Mosseri}}}, \ and\ \bibinfo {author} {\bibfnamefont {Patrick}\ \bibnamefont
			{{Butaud}}},\ }\bibfield  {title} {\enquote {\bibinfo {title} {{Interaction
					Induced Delocalization for Two Particles in a Periodic Potential}},}\ }\href
	{\doibase 10.1103/PhysRevLett.85.3906} {\bibfield  {journal} {\bibinfo
			{journal} {\prl}\ }\textbf {\bibinfo {volume} {85}},\ \bibinfo {pages}
		{3906--3909} (\bibinfo {year} {2000})},\ \Eprint
	{http://arxiv.org/abs/cond-mat/0005215} {arXiv:cond-mat/0005215
		[cond-mat.str-el]} \BibitemShut {NoStop}%
	\bibitem [{\citenamefont {{Vidal}}\ \emph {et~al.}(1998)\citenamefont
		{{Vidal}}, \citenamefont {{Mosseri}},\ and\ \citenamefont
		{{Dou{\c{c}}ot}}}]{FB-ABCage-1}%
	\BibitemOpen
	\bibfield  {author} {\bibinfo {author} {\bibfnamefont {Julien}\ \bibnamefont
			{{Vidal}}}, \bibinfo {author} {\bibfnamefont {R{\'e}my}\ \bibnamefont
			{{Mosseri}}}, \ and\ \bibinfo {author} {\bibfnamefont {Benoit}\ \bibnamefont
			{{Dou{\c{c}}ot}}},\ }\bibfield  {title} {\enquote {\bibinfo {title}
			{{Aharonov-Bohm Cages in Two-Dimensional Structures}},}\ }\href {\doibase
		10.1103/PhysRevLett.81.5888} {\bibfield  {journal} {\bibinfo  {journal}
			{\prl}\ }\textbf {\bibinfo {volume} {81}},\ \bibinfo {pages} {5888--5891}
		(\bibinfo {year} {1998})},\ \Eprint {http://arxiv.org/abs/cond-mat/9806068}
	{arXiv:cond-mat/9806068 [cond-mat.mes-hall]} \BibitemShut {NoStop}%
	\bibitem [{\citenamefont {{Landauer}}(1970)}]{Landauer-1D}%
	\BibitemOpen
	\bibfield  {author} {\bibinfo {author} {\bibfnamefont {Rolf}\ \bibnamefont
			{{Landauer}}},\ }\bibfield  {title} {\enquote {\bibinfo {title} {{Electrical
					resistance of disordered one-dimensional lattices}},}\ }\href {\doibase
		10.1080/14786437008238472} {\bibfield  {journal} {\bibinfo  {journal}
			{Philosophical Magazine}\ }\textbf {\bibinfo {volume} {21}},\ \bibinfo
		{pages} {863--867} (\bibinfo {year} {1970})}\BibitemShut {NoStop}%
	\bibitem [{\citenamefont {{B{\"u}ttiker}}(1986)}]{Buttiker-1}%
	\BibitemOpen
	\bibfield  {author} {\bibinfo {author} {\bibfnamefont {M.}~\bibnamefont
			{{B{\"u}ttiker}}},\ }\bibfield  {title} {\enquote {\bibinfo {title}
			{{Four-terminal phase-coherent conductance}},}\ }\href {\doibase
		10.1103/PhysRevLett.57.1761} {\bibfield  {journal} {\bibinfo  {journal}
			{\prl}\ }\textbf {\bibinfo {volume} {57}},\ \bibinfo {pages} {1761--1764}
		(\bibinfo {year} {1986})}\BibitemShut {NoStop}%
	\bibitem [{\citenamefont {{B{\"u}ttiker}}(1988)}]{Buttiker-2}%
	\BibitemOpen
	\bibfield  {author} {\bibinfo {author} {\bibfnamefont {M.}~\bibnamefont
			{{B{\"u}ttiker}}},\ }\bibfield  {title} {\enquote {\bibinfo {title} {{Absence
					of backscattering in the quantum Hall effect in multiprobe conductors}},}\
	}\href {\doibase 10.1103/PhysRevB.38.9375} {\bibfield  {journal} {\bibinfo
			{journal} {\prb}\ }\textbf {\bibinfo {volume} {38}},\ \bibinfo {pages}
		{9375--9389} (\bibinfo {year} {1988})}\BibitemShut {NoStop}%
	\bibitem [{\citenamefont {{de Picciotto}}\ \emph {et~al.}(2001)\citenamefont
		{{de Picciotto}}, \citenamefont {{Stormer}}, \citenamefont {{Pfeiffer}},
		\citenamefont {{Baldwin}},\ and\ \citenamefont {{West}}}]{FourTermMeasure-2}%
	\BibitemOpen
	\bibfield  {author} {\bibinfo {author} {\bibfnamefont {R.}~\bibnamefont {{de
					Picciotto}}}, \bibinfo {author} {\bibfnamefont {H.~L.}\ \bibnamefont
			{{Stormer}}}, \bibinfo {author} {\bibfnamefont {L.~N.}\ \bibnamefont
			{{Pfeiffer}}}, \bibinfo {author} {\bibfnamefont {K.~W.}\ \bibnamefont
			{{Baldwin}}}, \ and\ \bibinfo {author} {\bibfnamefont {K.~W.}\ \bibnamefont
			{{West}}},\ }\bibfield  {title} {\enquote {\bibinfo {title} {{Four-terminal
					resistance of a ballistic quantum wire}},}\ }\href {\doibase
		10.1038/35075009} {\bibfield  {journal} {\bibinfo  {journal} {\nat}\ }\textbf
		{\bibinfo {volume} {411}},\ \bibinfo {pages} {51--54} (\bibinfo {year}
		{2001})}\BibitemShut {NoStop}%
	\bibitem [{\citenamefont {{Jiang}}\ \emph {et~al.}(2014)\citenamefont
		{{Jiang}}, \citenamefont {{Liu}}, \citenamefont {{Feng}}, \citenamefont
		{{Sun}},\ and\ \citenamefont {{Xie}}}]{FourTermMeasure1}%
	\BibitemOpen
	\bibfield  {author} {\bibinfo {author} {\bibfnamefont {Hua}\ \bibnamefont
			{{Jiang}}}, \bibinfo {author} {\bibfnamefont {Haiwen}\ \bibnamefont {{Liu}}},
		\bibinfo {author} {\bibfnamefont {Ji}~\bibnamefont {{Feng}}}, \bibinfo
		{author} {\bibfnamefont {Qingfeng}\ \bibnamefont {{Sun}}}, \ and\ \bibinfo
		{author} {\bibfnamefont {X.~C.}\ \bibnamefont {{Xie}}},\ }\bibfield  {title}
	{\enquote {\bibinfo {title} {{Transport Discovery of Emerging Robust Helical
					Surface States in Z$_{2}$=0 Systems}},}\ }\href {\doibase
		10.1103/PhysRevLett.112.176601} {\bibfield  {journal} {\bibinfo  {journal}
			{\prl}\ }\textbf {\bibinfo {volume} {112}},\ \bibinfo {eid} {176601}
		(\bibinfo {year} {2014})},\ \Eprint {http://arxiv.org/abs/1403.3743}
	{arXiv:1403.3743 [cond-mat.mes-hall]} \BibitemShut {NoStop}%
	\bibitem [{\citenamefont {{Cook}}\ \emph {et~al.}(2011)\citenamefont {{Cook}},
		\citenamefont {{Dignard}},\ and\ \citenamefont {{Varga}}}]{Multiterminal-1}%
	\BibitemOpen
	\bibfield  {author} {\bibinfo {author} {\bibfnamefont {Brandon~G.}\
			\bibnamefont {{Cook}}}, \bibinfo {author} {\bibfnamefont {Peter}\
			\bibnamefont {{Dignard}}}, \ and\ \bibinfo {author} {\bibfnamefont
			{K{\'a}lm{\'a}n}\ \bibnamefont {{Varga}}},\ }\bibfield  {title} {\enquote
		{\bibinfo {title} {{Calculation of electron transport in multiterminal
					systems using complex absorbing potentials}},}\ }\href {\doibase
		10.1103/PhysRevB.83.205105} {\bibfield  {journal} {\bibinfo  {journal}
			{\prb}\ }\textbf {\bibinfo {volume} {83}},\ \bibinfo {eid} {205105} (\bibinfo
		{year} {2011})}\BibitemShut {NoStop}%
	\bibitem [{\citenamefont {{Hou}}\ \emph {et~al.}(2024)\citenamefont {{Hou}},
		\citenamefont {{Hu}},\ and\ \citenamefont {{Yang}}}]{Moire_3}%
	\BibitemOpen
	\bibfield  {author} {\bibinfo {author} {\bibfnamefont {Zhe}\ \bibnamefont
			{{Hou}}}, \bibinfo {author} {\bibfnamefont {Ya-Yun}\ \bibnamefont {{Hu}}}, \
		and\ \bibinfo {author} {\bibfnamefont {Guang-Wen}\ \bibnamefont {{Yang}}},\
	}\bibfield  {title} {\enquote {\bibinfo {title} {{Moir{\'e} pattern assisted
					commensuration resonance in disordered twisted bilayer graphene}},}\ }\href
	{\doibase 10.1103/PhysRevB.109.085412} {\bibfield  {journal} {\bibinfo
			{journal} {\prb}\ }\textbf {\bibinfo {volume} {109}},\ \bibinfo {eid}
		{085412} (\bibinfo {year} {2024})},\ \Eprint
	{http://arxiv.org/abs/2307.09587} {arXiv:2307.09587 [cond-mat.mes-hall]}
	\BibitemShut {NoStop}%
	\bibitem [{\citenamefont {{Guo}}\ \emph {et~al.}(2024)\citenamefont {{Guo}},
		\citenamefont {{Ma}}, \citenamefont {{Ying}},\ and\ \citenamefont
		{{Law}}}]{Maj_1}%
	\BibitemOpen
	\bibfield  {author} {\bibinfo {author} {\bibfnamefont {Xingyao}\ \bibnamefont
			{{Guo}}}, \bibinfo {author} {\bibfnamefont {Xinglei}\ \bibnamefont {{Ma}}},
		\bibinfo {author} {\bibfnamefont {Xuzhe}\ \bibnamefont {{Ying}}}, \ and\
		\bibinfo {author} {\bibfnamefont {K.~T.}\ \bibnamefont {{Law}}},\ }\bibfield
	{title} {\enquote {\bibinfo {title} {{Majorana Zero Modes in Lieb-Kitaev
					Model with Tunable Quantum Metric}},}\ }\href {\doibase
		10.48550/arXiv.2406.05789} {\bibfield  {journal} {\bibinfo  {journal} {arXiv
				e-prints}\ ,\ \bibinfo {eid} {arXiv:2406.05789}} (\bibinfo {year} {2024})},\
	\Eprint {http://arxiv.org/abs/2406.05789} {arXiv:2406.05789
		[cond-mat.supr-con]} \BibitemShut {NoStop}%
	\bibitem [{\citenamefont {{Li}}\ \emph {et~al.}(2024)\citenamefont {{Li}},
		\citenamefont {{Deng}}, \citenamefont {{Chen}}, \citenamefont {{Efetov}},\
		and\ \citenamefont {{Law}}}]{FB-JJ-3}%
	\BibitemOpen
	\bibfield  {author} {\bibinfo {author} {\bibfnamefont {Zhong C.~F.}\
			\bibnamefont {{Li}}}, \bibinfo {author} {\bibfnamefont {Yuxuan}\ \bibnamefont
			{{Deng}}}, \bibinfo {author} {\bibfnamefont {Shuai~A.}\ \bibnamefont
			{{Chen}}}, \bibinfo {author} {\bibfnamefont {Dmitri~K.}\ \bibnamefont
			{{Efetov}}}, \ and\ \bibinfo {author} {\bibfnamefont {K.~T.}\ \bibnamefont
			{{Law}}},\ }\bibfield  {title} {\enquote {\bibinfo {title} {{Flat Band
					Josephson Junctions with Quantum Metric}},}\ }\href {\doibase
		10.48550/arXiv.2404.09211} {\bibfield  {journal} {\bibinfo  {journal} {arXiv
				e-prints}\ ,\ \bibinfo {eid} {arXiv:2404.09211}} (\bibinfo {year} {2024})},\
	\Eprint {http://arxiv.org/abs/2404.09211} {arXiv:2404.09211
		[cond-mat.supr-con]} \BibitemShut {NoStop}%
	\bibitem [{sup()}]{supple}%
	\BibitemOpen
	\href@noop {} {}\bibinfo {note} {See Supplemental Material: I. Quantum
		geometry of the Lieb lattice; II. Interface state wave functions and the
		transmission; III. Green’s function calculations; IV. Kubo-Greenwood
		formula and conductivity in the clean limit; V. Diagrammatic calculations in
		the presence of disorder; VI. Numerical approaches and results}\BibitemShut
	{NoStop}%
	\bibitem [{\citenamefont {{Feilhauer}}\ and\ \citenamefont
		{{Mo{\v{s}}ko}}(2011)}]{Q1D-Diffu}%
	\BibitemOpen
	\bibfield  {author} {\bibinfo {author} {\bibfnamefont {J.}~\bibnamefont
			{{Feilhauer}}}\ and\ \bibinfo {author} {\bibfnamefont {M.}~\bibnamefont
			{{Mo{\v{s}}ko}}},\ }\bibfield  {title} {\enquote {\bibinfo {title} {{Quantum
					and Boltzmann transport in a quasi-one-dimensional wire with rough edges}},}\
	}\href {\doibase 10.1103/PhysRevB.83.245328} {\bibfield  {journal} {\bibinfo
			{journal} {\prb}\ }\textbf {\bibinfo {volume} {83}},\ \bibinfo {eid} {245328}
		(\bibinfo {year} {2011})},\ \Eprint {http://arxiv.org/abs/1011.6193}
	{arXiv:1011.6193 [cond-mat.mes-hall]} \BibitemShut {NoStop}%
	\bibitem [{\citenamefont {{Abrahams}}\ \emph {et~al.}(1979)\citenamefont
		{{Abrahams}}, \citenamefont {{Anderson}}, \citenamefont {{Licciardello}},\
		and\ \citenamefont {{Ramakrishnan}}}]{Dis-Cond-1}%
	\BibitemOpen
	\bibfield  {author} {\bibinfo {author} {\bibfnamefont {E.}~\bibnamefont
			{{Abrahams}}}, \bibinfo {author} {\bibfnamefont {P.~W.}\ \bibnamefont
			{{Anderson}}}, \bibinfo {author} {\bibfnamefont {D.~C.}\ \bibnamefont
			{{Licciardello}}}, \ and\ \bibinfo {author} {\bibfnamefont {T.~V.}\
			\bibnamefont {{Ramakrishnan}}},\ }\bibfield  {title} {\enquote {\bibinfo
			{title} {{Scaling Theory of Localization: Absence of Quantum Diffusion in Two
					Dimensions}},}\ }\href {\doibase 10.1103/PhysRevLett.42.673} {\bibfield
		{journal} {\bibinfo  {journal} {\prl}\ }\textbf {\bibinfo {volume} {42}},\
		\bibinfo {pages} {673--676} (\bibinfo {year} {1979})}\BibitemShut {NoStop}%
	\bibitem [{\citenamefont {{Mao}}\ \emph {et~al.}(2024)\citenamefont {{Mao}},
		\citenamefont {{Zeng}}, \citenamefont {{Shi}}, \citenamefont {{Wu}},
		\citenamefont {{Xie}}, \citenamefont {{Yuan}}, \citenamefont {{Zhang}},
		\citenamefont {{Dai}}, \citenamefont {{Chen}},\ and\ \citenamefont
		{{Pan}}}]{And-Loc-1}%
	\BibitemOpen
	\bibfield  {author} {\bibinfo {author} {\bibfnamefont {Yi-Yi}\ \bibnamefont
			{{Mao}}}, \bibinfo {author} {\bibfnamefont {Chao}\ \bibnamefont {{Zeng}}},
		\bibinfo {author} {\bibfnamefont {Yue-Ran}\ \bibnamefont {{Shi}}}, \bibinfo
		{author} {\bibfnamefont {Fei-Fei}\ \bibnamefont {{Wu}}}, \bibinfo {author}
		{\bibfnamefont {Yan-Jun}\ \bibnamefont {{Xie}}}, \bibinfo {author}
		{\bibfnamefont {Tao}\ \bibnamefont {{Yuan}}}, \bibinfo {author}
		{\bibfnamefont {Wei}\ \bibnamefont {{Zhang}}}, \bibinfo {author}
		{\bibfnamefont {Han-Ning}\ \bibnamefont {{Dai}}}, \bibinfo {author}
		{\bibfnamefont {Yu-Ao}\ \bibnamefont {{Chen}}}, \ and\ \bibinfo {author}
		{\bibfnamefont {Jian-Wei}\ \bibnamefont {{Pan}}},\ }\bibfield  {title}
	{\enquote {\bibinfo {title} {{Transition from flat-band localization to
					Anderson localization: Realization and characterization in a one-dimensional
					momentum lattice}},}\ }\href {\doibase 10.1103/PhysRevA.109.023304}
	{\bibfield  {journal} {\bibinfo  {journal} {\pra}\ }\textbf {\bibinfo
			{volume} {109}},\ \bibinfo {eid} {023304} (\bibinfo {year}
		{2024})}\BibitemShut {NoStop}%
	\bibitem [{\citenamefont {{Zeng}}\ \emph {et~al.}(2024)\citenamefont {{Zeng}},
		\citenamefont {{Shi}}, \citenamefont {{Mao}}, \citenamefont {{Wu}},
		\citenamefont {{Xie}}, \citenamefont {{Yuan}}, \citenamefont {{Zhang}},
		\citenamefont {{Dai}}, \citenamefont {{Chen}},\ and\ \citenamefont
		{{Pan}}}]{FBL-AL-1D-Exper}%
	\BibitemOpen
	\bibfield  {author} {\bibinfo {author} {\bibfnamefont {Chao}\ \bibnamefont
			{{Zeng}}}, \bibinfo {author} {\bibfnamefont {Yue-Ran}\ \bibnamefont {{Shi}}},
		\bibinfo {author} {\bibfnamefont {Yi-Yi}\ \bibnamefont {{Mao}}}, \bibinfo
		{author} {\bibfnamefont {Fei-Fei}\ \bibnamefont {{Wu}}}, \bibinfo {author}
		{\bibfnamefont {Yan-Jun}\ \bibnamefont {{Xie}}}, \bibinfo {author}
		{\bibfnamefont {Tao}\ \bibnamefont {{Yuan}}}, \bibinfo {author}
		{\bibfnamefont {Wei}\ \bibnamefont {{Zhang}}}, \bibinfo {author}
		{\bibfnamefont {Han-Ning}\ \bibnamefont {{Dai}}}, \bibinfo {author}
		{\bibfnamefont {Yu-Ao}\ \bibnamefont {{Chen}}}, \ and\ \bibinfo {author}
		{\bibfnamefont {Jian-Wei}\ \bibnamefont {{Pan}}},\ }\bibfield  {title}
	{\enquote {\bibinfo {title} {{Transition from Flat-Band Localization to
					Anderson Localization in a One-Dimensional Tasaki Lattice}},}\ }\href
	{\doibase 10.1103/PhysRevLett.132.063401} {\bibfield  {journal} {\bibinfo
			{journal} {\prl}\ }\textbf {\bibinfo {volume} {132}},\ \bibinfo {eid}
		{063401} (\bibinfo {year} {2024})}\BibitemShut {NoStop}%
	\bibitem [{\citenamefont {{H{\"o}rmann}}\ and\ \citenamefont
		{{Schmidt}}(2020)}]{FB-Loc-5}%
	\BibitemOpen
	\bibfield  {author} {\bibinfo {author} {\bibfnamefont {Max}\ \bibnamefont
			{{H{\"o}rmann}}}\ and\ \bibinfo {author} {\bibfnamefont {Kai~Phillip}\
			\bibnamefont {{Schmidt}}},\ }\bibfield  {title} {\enquote {\bibinfo {title}
			{{Dynamic structure factor of Heisenberg bilayer dimer phases in the presence
					of quenched disorder and frustration}},}\ }\href {\doibase
		10.1103/PhysRevB.102.094427} {\bibfield  {journal} {\bibinfo  {journal}
			{\prb}\ }\textbf {\bibinfo {volume} {102}},\ \bibinfo {eid} {094427}
		(\bibinfo {year} {2020})},\ \Eprint {http://arxiv.org/abs/2004.00565}
	{arXiv:2004.00565 [cond-mat.str-el]} \BibitemShut {NoStop}%
	\bibitem [{\citenamefont {{Fan}}\ \emph {et~al.}(2021)\citenamefont {{Fan}},
		\citenamefont {{Garcia}}, \citenamefont {{Cummings}}, \citenamefont
		{{Barrios-Vargas}}, \citenamefont {{Panhans}}, \citenamefont {{Harju}},
		\citenamefont {{Ortmann}},\ and\ \citenamefont {{Roche}}}]{Diff-Len-1}%
	\BibitemOpen
	\bibfield  {author} {\bibinfo {author} {\bibfnamefont {Zheyong}\ \bibnamefont
			{{Fan}}}, \bibinfo {author} {\bibfnamefont {Jos{\'e}~H.}\ \bibnamefont
			{{Garcia}}}, \bibinfo {author} {\bibfnamefont {Aron~W.}\ \bibnamefont
			{{Cummings}}}, \bibinfo {author} {\bibfnamefont {Jose~Eduardo}\ \bibnamefont
			{{Barrios-Vargas}}}, \bibinfo {author} {\bibfnamefont {Michel}\ \bibnamefont
			{{Panhans}}}, \bibinfo {author} {\bibfnamefont {Ari}\ \bibnamefont
			{{Harju}}}, \bibinfo {author} {\bibfnamefont {Frank}\ \bibnamefont
			{{Ortmann}}}, \ and\ \bibinfo {author} {\bibfnamefont {Stephan}\ \bibnamefont
			{{Roche}}},\ }\bibfield  {title} {\enquote {\bibinfo {title} {{Linear scaling
					quantum transport methodologies}},}\ }\href {\doibase
		10.1016/j.physrep.2020.12.001} {\bibfield  {journal} {\bibinfo  {journal}
			{Physics Reports}\ }\textbf {\bibinfo {volume} {903}},\ \bibinfo {pages}
		{1--69} (\bibinfo {year} {2021})},\ \Eprint {http://arxiv.org/abs/1811.07387}
	{arXiv:1811.07387 [cond-mat.mes-hall]} \BibitemShut {NoStop}%
	\bibitem [{\citenamefont {{Beenakker}}\ and\ \citenamefont {{van
				Houten}}(1991)}]{Review-QuantTrans}%
	\BibitemOpen
	\bibfield  {author} {\bibinfo {author} {\bibfnamefont {C.~W.~J.}\
			\bibnamefont {{Beenakker}}}\ and\ \bibinfo {author} {\bibfnamefont
			{H.}~\bibnamefont {{van Houten}}},\ }\bibfield  {title} {\enquote {\bibinfo
			{title} {{Quantum Transport in Semiconductor Nanostructures}},}\ }\href
	{\doibase 10.1016/S0081-1947(08)60091-0} {\bibfield  {journal} {\bibinfo
			{journal} {Solid State Physics}\ }\textbf {\bibinfo {volume} {44}},\ \bibinfo
		{pages} {1--228} (\bibinfo {year} {1991})}\BibitemShut {NoStop}%
	\bibitem [{\citenamefont {{Daumann}}\ and\ \citenamefont
		{{Dahm}}(2024)}]{Diffusion-MSD-1}%
	\BibitemOpen
	\bibfield  {author} {\bibinfo {author} {\bibfnamefont {Mirko}\ \bibnamefont
			{{Daumann}}}\ and\ \bibinfo {author} {\bibfnamefont {Thomas}\ \bibnamefont
			{{Dahm}}},\ }\bibfield  {title} {\enquote {\bibinfo {title} {{Anomalous
					diffusion, prethermalization, and particle binding in an interacting flat
					band system}},}\ }\href {\doibase 10.1088/1367-2630/ad4e5d} {\bibfield
		{journal} {\bibinfo  {journal} {New Journal of Physics}\ }\textbf {\bibinfo
			{volume} {26}},\ \bibinfo {eid} {063001} (\bibinfo {year} {2024})},\ \Eprint
	{http://arxiv.org/abs/2402.12180} {arXiv:2402.12180 [cond-mat.stat-mech]}
	\BibitemShut {NoStop}%
	\bibitem [{\citenamefont {{Markussen}}\ \emph {et~al.}(2006)\citenamefont
		{{Markussen}}, \citenamefont {{Rurali}}, \citenamefont {{Brandbyge}},\ and\
		\citenamefont {{Jauho}}}]{Diffusion-MSD-2}%
	\BibitemOpen
	\bibfield  {author} {\bibinfo {author} {\bibfnamefont {Troels}\ \bibnamefont
			{{Markussen}}}, \bibinfo {author} {\bibfnamefont {Riccardo}\ \bibnamefont
			{{Rurali}}}, \bibinfo {author} {\bibfnamefont {Mads}\ \bibnamefont
			{{Brandbyge}}}, \ and\ \bibinfo {author} {\bibfnamefont {Antti-Pekka}\
			\bibnamefont {{Jauho}}},\ }\bibfield  {title} {\enquote {\bibinfo {title}
			{{Electronic transport through Si nanowires: Role of bulk and surface
					disorder}},}\ }\href {\doibase 10.1103/PhysRevB.74.245313} {\bibfield
		{journal} {\bibinfo  {journal} {\prb}\ }\textbf {\bibinfo {volume} {74}},\
		\bibinfo {eid} {245313} (\bibinfo {year} {2006})},\ \Eprint
	{http://arxiv.org/abs/cond-mat/0606600} {arXiv:cond-mat/0606600
		[cond-mat.mes-hall]} \BibitemShut {NoStop}%
	\bibitem [{\citenamefont {{Zuo}}\ \emph {et~al.}(2024)\citenamefont {{Zuo}},
		\citenamefont {{Lin}},\ and\ \citenamefont {{Kang}}}]{FB-WavePacket-1}%
	\BibitemOpen
	\bibfield  {author} {\bibinfo {author} {\bibfnamefont {Zheng-Wei}\
			\bibnamefont {{Zuo}}}, \bibinfo {author} {\bibfnamefont {Jing-Run}\
			\bibnamefont {{Lin}}}, \ and\ \bibinfo {author} {\bibfnamefont {Dawei}\
			\bibnamefont {{Kang}}},\ }\bibfield  {title} {\enquote {\bibinfo {title}
			{{Topological inverse Anderson insulator}},}\ }\href {\doibase
		10.1103/PhysRevB.110.085157} {\bibfield  {journal} {\bibinfo  {journal}
			{\prb}\ }\textbf {\bibinfo {volume} {110}},\ \bibinfo {eid} {085157}
		(\bibinfo {year} {2024})},\ \Eprint {http://arxiv.org/abs/2408.15826}
	{arXiv:2408.15826 [cond-mat.mes-hall]} \BibitemShut {NoStop}%
	\bibitem [{\citenamefont {{Longhi}}(2021)}]{FB-Photon-1}%
	\BibitemOpen
	\bibfield  {author} {\bibinfo {author} {\bibfnamefont {Stefano}\ \bibnamefont
			{{Longhi}}},\ }\bibfield  {title} {\enquote {\bibinfo {title} {{Inverse
					Anderson transition in photonic cages}},}\ }\href {\doibase
		10.1364/OL.430196} {\bibfield  {journal} {\bibinfo  {journal} {Optics
				Letters}\ }\textbf {\bibinfo {volume} {46}},\ \bibinfo {pages} {2872}
		(\bibinfo {year} {2021})},\ \Eprint {http://arxiv.org/abs/2106.00231}
	{arXiv:2106.00231 [physics.optics]} \BibitemShut {NoStop}%
	\bibitem [{\citenamefont {{Rosen}}\ \emph {et~al.}(2025)\citenamefont
		{{Rosen}}, \citenamefont {{Muschinske}}, \citenamefont {{Barrett}},
		\citenamefont {{Rower}}, \citenamefont {{Das}}, \citenamefont {{Kim}},
		\citenamefont {{Niedzielski}}, \citenamefont {{Schuldt}}, \citenamefont
		{{Serniak}}, \citenamefont {{Schwartz}}, \citenamefont {{Yoder}},
		\citenamefont {{Grover}},\ and\ \citenamefont {{Oliver}}}]{FB-SC-Qbit-Array}%
	\BibitemOpen
	\bibfield  {author} {\bibinfo {author} {\bibfnamefont {Ilan~T.}\ \bibnamefont
			{{Rosen}}}, \bibinfo {author} {\bibfnamefont {Sarah}\ \bibnamefont
			{{Muschinske}}}, \bibinfo {author} {\bibfnamefont {Cora~N.}\ \bibnamefont
			{{Barrett}}}, \bibinfo {author} {\bibfnamefont {David~A.}\ \bibnamefont
			{{Rower}}}, \bibinfo {author} {\bibfnamefont {Rabindra}\ \bibnamefont
			{{Das}}}, \bibinfo {author} {\bibfnamefont {David~K.}\ \bibnamefont {{Kim}}},
		\bibinfo {author} {\bibfnamefont {Bethany~M.}\ \bibnamefont {{Niedzielski}}},
		\bibinfo {author} {\bibfnamefont {Meghan}\ \bibnamefont {{Schuldt}}},
		\bibinfo {author} {\bibfnamefont {Kyle}\ \bibnamefont {{Serniak}}}, \bibinfo
		{author} {\bibfnamefont {Mollie~E.}\ \bibnamefont {{Schwartz}}}, \bibinfo
		{author} {\bibfnamefont {Jonilyn~L.}\ \bibnamefont {{Yoder}}}, \bibinfo
		{author} {\bibfnamefont {Jeffrey~A.}\ \bibnamefont {{Grover}}}, \ and\
		\bibinfo {author} {\bibfnamefont {William~D.}\ \bibnamefont {{Oliver}}},\
	}\bibfield  {title} {\enquote {\bibinfo {title} {{Flat-Band (De)localization
					Emulated with a Superconducting Qubit Array}},}\ }\href {\doibase
		10.1103/PhysRevX.15.021091} {\bibfield  {journal} {\bibinfo  {journal}
			{Physical Review X}\ }\textbf {\bibinfo {volume} {15}},\ \bibinfo {eid}
		{021091} (\bibinfo {year} {2025})},\ \Eprint
	{http://arxiv.org/abs/2410.07878} {arXiv:2410.07878 [cond-mat.mes-hall]}
	\BibitemShut {NoStop}%
	\bibitem [{\citenamefont {{Leykam}}\ \emph {et~al.}(2013)\citenamefont
		{{Leykam}}, \citenamefont {{Flach}}, \citenamefont {{Bahat-Treidel}},\ and\
		\citenamefont {{Desyatnikov}}}]{DL-Diffusion}%
	\BibitemOpen
	\bibfield  {author} {\bibinfo {author} {\bibfnamefont {Daniel}\ \bibnamefont
			{{Leykam}}}, \bibinfo {author} {\bibfnamefont {Sergej}\ \bibnamefont
			{{Flach}}}, \bibinfo {author} {\bibfnamefont {Omri}\ \bibnamefont
			{{Bahat-Treidel}}}, \ and\ \bibinfo {author} {\bibfnamefont {Anton~S.}\
			\bibnamefont {{Desyatnikov}}},\ }\bibfield  {title} {\enquote {\bibinfo
			{title} {{Flat band states: Disorder and nonlinearity}},}\ }\href {\doibase
		10.1103/PhysRevB.88.224203} {\bibfield  {journal} {\bibinfo  {journal}
			{\prb}\ }\textbf {\bibinfo {volume} {88}},\ \bibinfo {eid} {224203} (\bibinfo
		{year} {2013})},\ \Eprint {http://arxiv.org/abs/1305.7287} {arXiv:1305.7287
		[cond-mat.dis-nn]} \BibitemShut {NoStop}%
	\bibitem [{\citenamefont {{Kawa}}\ and\ \citenamefont
		{{Machnikowski}}(2020)}]{Diffusion-MSD-3}%
	\BibitemOpen
	\bibfield  {author} {\bibinfo {author} {\bibfnamefont {K.}~\bibnamefont
			{{Kawa}}}\ and\ \bibinfo {author} {\bibfnamefont {P.}~\bibnamefont
			{{Machnikowski}}},\ }\bibfield  {title} {\enquote {\bibinfo {title}
			{{Diffusion of excitations and power-law localization in strongly disordered
					systems with long-range coupling}},}\ }\href {\doibase
		10.1103/PhysRevB.102.174203} {\bibfield  {journal} {\bibinfo  {journal}
			{\prb}\ }\textbf {\bibinfo {volume} {102}},\ \bibinfo {eid} {174203}
		(\bibinfo {year} {2020})}\BibitemShut {NoStop}%
	\bibitem [{\citenamefont {Tutunnikov}\ \emph {et~al.}(2023)\citenamefont
		{Tutunnikov}, \citenamefont {Chuang},\ and\ \citenamefont
		{Cao}}]{Diffusion-Noise-1}%
	\BibitemOpen
	\bibfield  {author} {\bibinfo {author} {\bibfnamefont {Ilia}\ \bibnamefont
			{Tutunnikov}}, \bibinfo {author} {\bibfnamefont {Chern}\ \bibnamefont
			{Chuang}}, \ and\ \bibinfo {author} {\bibfnamefont {Jianshu}\ \bibnamefont
			{Cao}},\ }\bibfield  {title} {\enquote {\bibinfo {title} {Coherent spatial
				control of wave packet dynamics on quantum lattices},}\ }\href {\doibase
		10.1021/acs.jpclett.3c03047} {\bibfield  {journal} {\bibinfo  {journal} {The
				Journal of Physical Chemistry Letters}\ }\textbf {\bibinfo {volume} {14}},\
		\bibinfo {pages} {11632--11639} (\bibinfo {year} {2023})},\ \Eprint
	{http://arxiv.org/abs/https://doi.org/10.1021/acs.jpclett.3c03047}
	{https://doi.org/10.1021/acs.jpclett.3c03047} \BibitemShut {NoStop}%
	\bibitem [{\citenamefont {Akkermans}\ and\ \citenamefont
		{Montambaux}(2007)}]{Book-Meso}%
	\BibitemOpen
	\bibfield  {author} {\bibinfo {author} {\bibfnamefont {Eric}\ \bibnamefont
			{Akkermans}}\ and\ \bibinfo {author} {\bibfnamefont {Gilles}\ \bibnamefont
			{Montambaux}},\ }\href {\doibase 10.1017/CBO9780511618833} {\emph {\bibinfo
			{title} {Mesoscopic {{Physics}} of {{Electrons}} and {{Photons}}}}},\
	\bibinfo {edition} {1st}\ ed.\ (\bibinfo  {publisher} {Cambridge University
		Press},\ \bibinfo {year} {2007})\BibitemShut {NoStop}%
	\bibitem [{\citenamefont {Bruus}\ and\ \citenamefont
		{Flensberg}(2004)}]{Book-Bruus}%
	\BibitemOpen
	\bibfield  {author} {\bibinfo {author} {\bibfnamefont {Henrik}\ \bibnamefont
			{Bruus}}\ and\ \bibinfo {author} {\bibfnamefont {Karsten}\ \bibnamefont
			{Flensberg}},\ }\href {\doibase 10.1093/oso/9780198566335.001.0001} {\emph
		{\bibinfo {title} {Many–Body Quantum Theory in Condensed Matter Physics: An
				Introduction}}}\ (\bibinfo  {publisher} {Oxford University Press},\ \bibinfo
	{year} {2004})\BibitemShut {NoStop}%
	\bibitem [{\citenamefont {Allen}(2006)}]{Book-allen}%
	\BibitemOpen
	\bibfield  {author} {\bibinfo {author} {\bibfnamefont {P.B.}\ \bibnamefont
			{Allen}},\ }\bibfield  {title} {\enquote {\bibinfo {title} {Chapter 6
				{{Electron Transport}}},}\ }in\ \href {\doibase
		10.1016/S1572-0934(06)02006-3} {\emph {\bibinfo {booktitle} {Contemporary
				{{Concepts}} of {{Condensed Matter Science}}}}},\ Vol.~\bibinfo {volume} {2}\
	(\bibinfo  {publisher} {Elsevier},\ \bibinfo {year} {2006})\ pp.\ \bibinfo
	{pages} {165--218}\BibitemShut {NoStop}%
	\bibitem [{\citenamefont {{Rhim}}\ and\ \citenamefont
		{{Yang}}(2019)}]{Flat-Tasaki-1}%
	\BibitemOpen
	\bibfield  {author} {\bibinfo {author} {\bibfnamefont {Jun-Won}\ \bibnamefont
			{{Rhim}}}\ and\ \bibinfo {author} {\bibfnamefont {Bohm-Jung}\ \bibnamefont
			{{Yang}}},\ }\bibfield  {title} {\enquote {\bibinfo {title} {{Classification
					of flat bands according to the band-crossing singularity of Bloch wave
					functions}},}\ }\href {\doibase 10.1103/PhysRevB.99.045107} {\bibfield
		{journal} {\bibinfo  {journal} {\prb}\ }\textbf {\bibinfo {volume} {99}},\
		\bibinfo {eid} {045107} (\bibinfo {year} {2019})},\ \Eprint
	{http://arxiv.org/abs/1808.05926} {arXiv:1808.05926 [cond-mat.str-el]}
	\BibitemShut {NoStop}%
	\bibitem [{\citenamefont {{Bistritzer}}\ and\ \citenamefont
		{{MacDonald}}(2011)}]{Moire_1}%
	\BibitemOpen
	\bibfield  {author} {\bibinfo {author} {\bibfnamefont {Rafi}\ \bibnamefont
			{{Bistritzer}}}\ and\ \bibinfo {author} {\bibfnamefont {Allan~H.}\
			\bibnamefont {{MacDonald}}},\ }\bibfield  {title} {\enquote {\bibinfo {title}
			{{Moir{\'e} bands in twisted double-layer graphene}},}\ }\href {\doibase
		10.1073/pnas.1108174108} {\bibfield  {journal} {\bibinfo  {journal}
			{Proceedings of the National Academy of Science}\ }\textbf {\bibinfo {volume}
			{108}},\ \bibinfo {pages} {12233--12237} (\bibinfo {year} {2011})},\ \Eprint
	{http://arxiv.org/abs/1009.4203} {arXiv:1009.4203 [cond-mat.mes-hall]}
	\BibitemShut {NoStop}%
	\bibitem [{\citenamefont {{Nuckolls}}\ and\ \citenamefont
		{{Yazdani}}(2024)}]{Moire_2}%
	\BibitemOpen
	\bibfield  {author} {\bibinfo {author} {\bibfnamefont {Kevin~P.}\
			\bibnamefont {{Nuckolls}}}\ and\ \bibinfo {author} {\bibfnamefont {Ali}\
			\bibnamefont {{Yazdani}}},\ }\bibfield  {title} {\enquote {\bibinfo {title}
			{{A Microscopic Perspective on Moir{\'e} Materials}},}\ }\href {\doibase
		10.48550/arXiv.2404.08044} {\bibfield  {journal} {\bibinfo  {journal} {arXiv
				e-prints}\ ,\ \bibinfo {eid} {arXiv:2404.08044}} (\bibinfo {year} {2024})},\
	\Eprint {http://arxiv.org/abs/2404.08044} {arXiv:2404.08044
		[cond-mat.mes-hall]} \BibitemShut {NoStop}%
	\bibitem [{\citenamefont {{Guerrero}}\ \emph {et~al.}(2025)\citenamefont
		{{Guerrero}}, \citenamefont {{Nguyen}}, \citenamefont {{Romeral}},
		\citenamefont {{Cummings}}, \citenamefont {{Garcia}}, \citenamefont
		{{Charlier}},\ and\ \citenamefont {{Roche}}}]{Dis-MATBG-1}%
	\BibitemOpen
	\bibfield  {author} {\bibinfo {author} {\bibfnamefont {Pedro~Alc{\'a}zar}\
			\bibnamefont {{Guerrero}}}, \bibinfo {author} {\bibfnamefont {Viet-Hung}\
			\bibnamefont {{Nguyen}}}, \bibinfo {author} {\bibfnamefont
			{Jorge~Mart{\'\i}nez}\ \bibnamefont {{Romeral}}}, \bibinfo {author}
		{\bibfnamefont {Aron~W.}\ \bibnamefont {{Cummings}}}, \bibinfo {author}
		{\bibfnamefont {Jos{\'e}-Hugo}\ \bibnamefont {{Garcia}}}, \bibinfo {author}
		{\bibfnamefont {Jean-Christophe}\ \bibnamefont {{Charlier}}}, \ and\ \bibinfo
		{author} {\bibfnamefont {Stephan}\ \bibnamefont {{Roche}}},\ }\bibfield
	{title} {\enquote {\bibinfo {title} {{Disorder-Induced Delocalization in
					Magic-Angle Twisted Bilayer Graphene}},}\ }\href {\doibase
		10.1103/PhysRevLett.134.126301} {\bibfield  {journal} {\bibinfo  {journal}
			{\prl}\ }\textbf {\bibinfo {volume} {134}},\ \bibinfo {eid} {126301}
		(\bibinfo {year} {2025})},\ \Eprint {http://arxiv.org/abs/2401.08265}
	{arXiv:2401.08265 [cond-mat.mes-hall]} \BibitemShut {NoStop}%
	\bibitem [{\citenamefont {{Chen}}\ \emph {et~al.}(2024)\citenamefont {{Chen}},
		\citenamefont {{Li}}, \citenamefont {{Xie}}, \citenamefont {{Zhang}},
		\citenamefont {{Lam}}, \citenamefont {{Tang}},\ and\ \citenamefont
		{{Lin}}}]{QDA}%
	\BibitemOpen
	\bibfield  {author} {\bibinfo {author} {\bibfnamefont {Cheng-Yi}\
			\bibnamefont {{Chen}}}, \bibinfo {author} {\bibfnamefont {En}~\bibnamefont
			{{Li}}}, \bibinfo {author} {\bibfnamefont {Huilin}\ \bibnamefont {{Xie}}},
		\bibinfo {author} {\bibfnamefont {Jianyu}\ \bibnamefont {{Zhang}}}, \bibinfo
		{author} {\bibfnamefont {Jacky Wing~Yip}\ \bibnamefont {{Lam}}}, \bibinfo
		{author} {\bibfnamefont {Ben~Zhong}\ \bibnamefont {{Tang}}}, \ and\ \bibinfo
		{author} {\bibfnamefont {Nian}\ \bibnamefont {{Lin}}},\ }\bibfield  {title}
	{\enquote {\bibinfo {title} {{Isolated flat band in artificially designed
					Lieb lattice based on macrocycle supramolecular crystal}},}\ }\href {\doibase
		10.1038/s43246-024-00501-8} {\bibfield  {journal} {\bibinfo  {journal}
			{Communications Materials}\ }\textbf {\bibinfo {volume} {5}},\ \bibinfo {eid}
		{54} (\bibinfo {year} {2024})}\BibitemShut {NoStop}%
	\bibitem [{\citenamefont {{Xia}}\ \emph {et~al.}(2016)\citenamefont {{Xia}},
		\citenamefont {{Hu}}, \citenamefont {{Song}}, \citenamefont {{Zong}},
		\citenamefont {{Tang}},\ and\ \citenamefont {{Chen}}}]{Op_Lat_1}%
	\BibitemOpen
	\bibfield  {author} {\bibinfo {author} {\bibfnamefont {Shiqiang}\
			\bibnamefont {{Xia}}}, \bibinfo {author} {\bibfnamefont {Yi}~\bibnamefont
			{{Hu}}}, \bibinfo {author} {\bibfnamefont {Daohong}\ \bibnamefont {{Song}}},
		\bibinfo {author} {\bibfnamefont {Yuanyuan}\ \bibnamefont {{Zong}}}, \bibinfo
		{author} {\bibfnamefont {Liqin}\ \bibnamefont {{Tang}}}, \ and\ \bibinfo
		{author} {\bibfnamefont {Zhigang}\ \bibnamefont {{Chen}}},\ }\bibfield
	{title} {\enquote {\bibinfo {title} {{Demonstration of flat-band image
					transmission in optically induced Lieb photonic lattices}},}\ }\href
	{\doibase 10.1364/OL.41.001435} {\bibfield  {journal} {\bibinfo  {journal}
			{Optics Letters}\ }\textbf {\bibinfo {volume} {41}},\ \bibinfo {pages} {1435}
		(\bibinfo {year} {2016})}\BibitemShut {NoStop}%
	\bibitem [{\citenamefont {{Kang}}\ \emph {et~al.}(2020)\citenamefont {{Kang}},
		\citenamefont {{Ye}}, \citenamefont {{Fang}}, \citenamefont {{You}},
		\citenamefont {{Levitan}}, \citenamefont {{Han}}, \citenamefont {{Facio}},
		\citenamefont {{Jozwiak}}, \citenamefont {{Bostwick}}, \citenamefont
		{{Rotenberg}}, \citenamefont {{Chan}}, \citenamefont {{McDonald}},
		\citenamefont {{Graf}}, \citenamefont {{Kaznatcheev}}, \citenamefont
		{{Vescovo}}, \citenamefont {{Bell}}, \citenamefont {{Kaxiras}}, \citenamefont
		{{van den Brink}}, \citenamefont {{Richter}}, \citenamefont {{Prasad
				Ghimire}}, \citenamefont {{Checkelsky}},\ and\ \citenamefont
		{{Comin}}}]{Kagome_1}%
	\BibitemOpen
	\bibfield  {author} {\bibinfo {author} {\bibfnamefont {Mingu}\ \bibnamefont
			{{Kang}}}, \bibinfo {author} {\bibfnamefont {Linda}\ \bibnamefont {{Ye}}},
		\bibinfo {author} {\bibfnamefont {Shiang}\ \bibnamefont {{Fang}}}, \bibinfo
		{author} {\bibfnamefont {Jhih-Shih}\ \bibnamefont {{You}}}, \bibinfo {author}
		{\bibfnamefont {Abe}\ \bibnamefont {{Levitan}}}, \bibinfo {author}
		{\bibfnamefont {Minyong}\ \bibnamefont {{Han}}}, \bibinfo {author}
		{\bibfnamefont {Jorge~I.}\ \bibnamefont {{Facio}}}, \bibinfo {author}
		{\bibfnamefont {Chris}\ \bibnamefont {{Jozwiak}}}, \bibinfo {author}
		{\bibfnamefont {Aaron}\ \bibnamefont {{Bostwick}}}, \bibinfo {author}
		{\bibfnamefont {Eli}\ \bibnamefont {{Rotenberg}}}, \bibinfo {author}
		{\bibfnamefont {Mun~K.}\ \bibnamefont {{Chan}}}, \bibinfo {author}
		{\bibfnamefont {Ross~D.}\ \bibnamefont {{McDonald}}}, \bibinfo {author}
		{\bibfnamefont {David}\ \bibnamefont {{Graf}}}, \bibinfo {author}
		{\bibfnamefont {Konstantine}\ \bibnamefont {{Kaznatcheev}}}, \bibinfo
		{author} {\bibfnamefont {Elio}\ \bibnamefont {{Vescovo}}}, \bibinfo {author}
		{\bibfnamefont {David~C.}\ \bibnamefont {{Bell}}}, \bibinfo {author}
		{\bibfnamefont {Efthimios}\ \bibnamefont {{Kaxiras}}}, \bibinfo {author}
		{\bibfnamefont {Jeroen}\ \bibnamefont {{van den Brink}}}, \bibinfo {author}
		{\bibfnamefont {Manuel}\ \bibnamefont {{Richter}}}, \bibinfo {author}
		{\bibfnamefont {Madhav}\ \bibnamefont {{Prasad Ghimire}}}, \bibinfo {author}
		{\bibfnamefont {Joseph~G.}\ \bibnamefont {{Checkelsky}}}, \ and\ \bibinfo
		{author} {\bibfnamefont {Riccardo}\ \bibnamefont {{Comin}}},\ }\bibfield
	{title} {\enquote {\bibinfo {title} {{Dirac fermions and flat bands in the
					ideal kagome metal FeSn}},}\ }\href {\doibase 10.1038/s41563-019-0531-0}
	{\bibfield  {journal} {\bibinfo  {journal} {Nature Materials}\ }\textbf
		{\bibinfo {volume} {19}},\ \bibinfo {pages} {163--169} (\bibinfo {year}
		{2020})},\ \Eprint {http://arxiv.org/abs/1906.02167} {arXiv:1906.02167
		[cond-mat.str-el]} \BibitemShut {NoStop}%
	\bibitem [{\citenamefont {{Chen}}\ \emph {et~al.}(2023)\citenamefont {{Chen}},
		\citenamefont {{Huang}}, \citenamefont {{Jiang}},\ and\ \citenamefont
		{{Hu}}}]{CLS-1}%
	\BibitemOpen
	\bibfield  {author} {\bibinfo {author} {\bibfnamefont {Yuge}\ \bibnamefont
			{{Chen}}}, \bibinfo {author} {\bibfnamefont {Juntao}\ \bibnamefont
			{{Huang}}}, \bibinfo {author} {\bibfnamefont {Kun}\ \bibnamefont {{Jiang}}},
		\ and\ \bibinfo {author} {\bibfnamefont {Jiangping}\ \bibnamefont {{Hu}}},\
	}\bibfield  {title} {\enquote {\bibinfo {title} {{Decoding flat bands from
					compact localized states}},}\ }\href {\doibase 10.1016/j.scib.2023.11.032}
	{\bibfield  {journal} {\bibinfo  {journal} {Science Bulletin}\ }\textbf
		{\bibinfo {volume} {68}},\ \bibinfo {pages} {3165--3171} (\bibinfo {year}
		{2023})},\ \Eprint {http://arxiv.org/abs/2212.13526} {arXiv:2212.13526
		[cond-mat.str-el]} \BibitemShut {NoStop}%
	\bibitem [{\citenamefont {{Leykam}}\ \emph {et~al.}(2017)\citenamefont
		{{Leykam}}, \citenamefont {{Bodyfelt}}, \citenamefont {{Desyatnikov}},\ and\
		\citenamefont {{Flach}}}]{CLS-2}%
	\BibitemOpen
	\bibfield  {author} {\bibinfo {author} {\bibfnamefont {Daniel}\ \bibnamefont
			{{Leykam}}}, \bibinfo {author} {\bibfnamefont {Joshua~D.}\ \bibnamefont
			{{Bodyfelt}}}, \bibinfo {author} {\bibfnamefont {Anton~S.}\ \bibnamefont
			{{Desyatnikov}}}, \ and\ \bibinfo {author} {\bibfnamefont {Sergej}\
			\bibnamefont {{Flach}}},\ }\bibfield  {title} {\enquote {\bibinfo {title}
			{{Localization of weakly disordered flat band states}},}\ }\href {\doibase
		10.1140/epjb/e2016-70551-2} {\bibfield  {journal} {\bibinfo  {journal}
			{European Physical Journal B}\ }\textbf {\bibinfo {volume} {90}},\ \bibinfo
		{eid} {1} (\bibinfo {year} {2017})},\ \Eprint
	{http://arxiv.org/abs/1601.03784} {arXiv:1601.03784 [cond-mat.dis-nn]}
	\BibitemShut {NoStop}%
\end{thebibliography}
%

\end{document}